\documentclass[prb,aps,twocolumn]{revtex4}

\usepackage{bm}
\usepackage{dcolumn}
\usepackage{graphicx}
\usepackage{amsfonts}
\usepackage{amsmath}
\usepackage{amssymb}
\usepackage{color}

\begin{document}

%\preprint{APS/123-QED}
\title{Correlations in Quantum Spin Ladders with Site and Bond Dilution}

\author{Kien Trinh}
\email{ktrinh@usc.edu}
\affiliation{Department of Physics and Astronomy, University of Southern California, Los Angeles, CA 90089, USA}
\author{Stephan Haas}
\affiliation{Department of Physics and Astronomy, University of Southern California, Los Angeles, CA 90089, USA}
\author{Rong Yu}
\affiliation{Department of Physics \& Astronomy, Rice University, Houston, TX 77005, USA}
\author{Tommaso Roscilde}
\affiliation{Laboratoire de Physique, CNRS UMR 5672, Ecole Normale Sup\'erieure de Lyon, Universit\'e de Lyon, 46 All\'ee d'Italie, Lyon, F-69364, France}

\date{\today}

\begin{abstract}
We investigate the effects of quenched disorder, in the form of site and bond dilution, on the physics of the $S=1/2$ antiferromagnetic Heisenberg model on even-leg ladders. Site dilution is found to prune rung singlets and thus create localized moments which interact via a random, unfrustrated network of effective couplings, realizing a random-exchange Heisenberg model (REHM) in one spatial dimension. This system exhibits a power-law diverging correlation length as the temperature decreases. Contrary to previous claims, we observe that the scaling exponent is non-universal, i.e., doping dependent. This finding can be explained by the discrete nature of the values taken by the effective exchange couplings in the doped ladder. Bond dilution on even-leg ladders leads to a more complex evolution with doping of correlations, which are weakly enhanced in 2-leg ladders, and are even suppressed for low dilution in the case of 4-leg and 6-leg ladders. We clarify the different aspects of correlation enhancement and suppression due to bond dilution by isolating the contributions of rung-bond dilution and leg-bond dilution.
\end{abstract}
\maketitle

\section{Introduction}

Low-dimensional quantum Heisenberg antiferromagnets are known to exhibit non-trivial properties due to quantum fluctuations. A most striking example comes from spin ladders: interpolating between the chain geometry and the square lattice geometry, they have provided an interesting new playground to explore the physics of low-dimensional strongly correlated electron systems \cite{DagottoR96}. Specifically, it has been shown that AF Heisenberg spin-$1/2$ ladders with an even number of legs are quantum spin liquids with purely short-range spin correlations. Their spin correlations decay exponentially due to a finite singlet-triplet gap in the energy spectrum. In contrast, Heisenberg ladders with an odd number of legs display properties similar to those of single chains at low temperature, namely gapless spin excitations, and power-law spin correlations. This difference between even-leg and odd-leg ladders was predicted by theory and confirmed by experiments in a variety of systems \cite{DagottoR96}.

In this paper we focus on disordered spin ladders. For the case of doping with static non-magnetic impurities, one distinguishes between site and bond disorder. Site dilution occurs upon doping the magnetic ions with non-magnetic ones (as for instance substitution of Cu$^{2+}$ with Zn$^{2+}$). This leads to the formation of local moments (LMs) close to the dopant site \cite{SigristF96, Sandviketal97, Martinsetal97}. Effective residual interactions between the LMs lead to long-range antiferromagnetic ordering\cite{Wessel01, Laflorencie03, Yasuda01, Imada97, Weber06}, experimentally detected in systems such as $\mathrm{Sr(Cu_{1-z}Zn_z)O_3}$ (Ref.~\onlinecite{Azumaetal97}) and Bi(Cu$_{\rm 1-z}$Zn$_{\rm z})_2$PO$_6$  (Ref.~\onlinecite{Bobroffetal09}). Bond disorder occurs instead when the dopant ions replace the ions which act as bridges between the magnetic ions. For instance, bond disorder can be introduced in IPA-$\mathrm{CuCl_3}$ by a partial substitution of non-magnetic $\mathrm{Br^-}$ for the likewise non-magnetic $\mathrm{Cl^-}$, affecting the bond angles in the $\mathrm{Cu}$-halogen-halogen-$\mathrm{Cu}$ superexchange pathways \cite{Manakaetal08, Manakaetal09, Hongetal10}. The reduced strength of magnetic interactions on the affected bonds leads to Bose-glass behavior when a strong magnetic field is applied.

From the theory side, a number of studies have addressed the effects of site and bond disorder on Heisenberg ladders. Ref.~\onlinecite{GrevenB98} examined the correlation length of randomly site-diluted spin-1/2 Heisenberg 2-leg ladders at weak and intermediate interchain couplings, showing an apparent divergence of the spin correlations at low temperatures, due to the presence of the impurities. A related divergence of the staggered susceptibility has been reported in Ref.~\onlinecite{Miyazakietal97}. This behavior has been related to that of the random-exchange Heisenberg model (REHM) \cite{Furusakietal94,Westerbergetal95}, describing the effective random interactions between LMs \cite{SigristF96, Nagaosaetal96}. In a more recent study, three of us have investigated bond-diluted 2-leg ladders \cite{Yuetal05, Yuetal06} and observed enhancement of spin correlations due to bond dilution. However, a general understanding of the mechanisms for enhancement of correlations in site- and bond-diluted Heisenberg ladders is still lacking. It is also unclear whether such an enhancement of spin correlations generally exists for even-leg ladders with a number of legs greater than two.

In this paper, we present a study of site- and bond-diluted spin-1/2 AF Heisenberg ladders with n=2, 4 and 6 legs. Making use of quantum Monte Carlo simulations, we can address the correlation length of the system down to extremely small temperatures at which the asymptotic $T\to 0$ behavior sets in. Our main findings can be summarized as follows:
\begin{enumerate}
\item For site-diluted even-leg ladders, we observe a strong enhancement of correlations upon doping up to $n=6$ legs. In the specific case of 2-leg ladders, we find that the system realizes the physics of the REHM with a \emph{discrete} distribution of the effective couplings, leading to a \emph{non-universal} behavior at low temperatures; hence the known predictions for the universal regime of the REHM with a continuous distribution of couplings are expected not to apply to realistic models of doped spin ladders.
\item For bond-diluted ladders, we find that correlations are \emph{suppressed} by a low level of dilution, due to dilution-induced dimerization. Hence the system is first driven to a gapless phase with short-range correlations. The correlation length becomes logarithmically divergent for vanishing temperatures only beyond a critical dilution.
\end{enumerate}

\section{Model and quantities of interest}

We study the Heisenberg model on $n$-leg ladders
\begin{eqnarray}
H&=& J_l \sum_{m=1}^n \sum_{i=1}^L p^{(l)}_{i,m} {{\bm S}_{i,m} \cdot {\bm S}_{i+1,m}} \nonumber \\
&+& J_r \sum_{m=1}^{n-1}
\sum_{i=1}^L p^{(r)}_{i,m} {{\bm S}_{i,m} \cdot {\bm S}_{i,m+1}}
\end{eqnarray}
where  ${\bm S}_{i,m}$  is the quantum spin operator at site $i=1,...,L$ along the $m$-th leg. The first term in the Hamiltonian describes interactions along the legs with coupling $J_l$, whereas the second term represents the rung interaction with coupling $J_r$. The random variables $p^{(l)}_{i,m}$, $p^{(r)}_{i,m}$ express the site or bond dilution: in the case of site dilution, $p^{(l)}_{i,m} = \epsilon_{i,m}  \epsilon_{i+1,m}$ and $p^{(r)}_{i,m} = \epsilon_{i,m}  \epsilon_{i,m+1}$, where $\epsilon_{i,m}$ takes value 1 if the site $(i,m)$ is occupied (with probability $1-z$) or 0 if it is empty (with probability $z$). In the case of bond dilution, $p^{(r)}_{i,m}=p^{(l)}_{i,m}/2$ take values 0 with probability $z/3$ if the bond is not occupied, and take values 1 otherwise. The simulations are performed using the Stochastic Series Expansion (SSE) quantum Monte Carlo (QMC) method based on the directed loop algorithm \cite{SyljuasenS02}. Periodic boundary conditions (PBC) along the leg direction are used. To access the regime of very low temperatures, at which the asymptotic low-$T$ behavior of the correlation length sets in, we have made use a $\beta$-doubling scheme \cite{Sandvik02}, allowing us to efficiently access inverse temperatures up to $\beta = 4096$.

The main objective of this paper is to study the correlation length along the leg direction, which is calculated via the disorder-averaged second-moment estimator:
\cite{Cooperetal82}:
\begin{equation}
\xi =\frac{L}{2\pi}\sqrt{\frac{[S(\pi,\pi)]_{\rm av}}{[S(\pi+2\pi/L,\pi)]_{\rm av}}-1} .
\label{e.2ndmoment}
\end{equation}
Here $S(\bm q)$ is the time-averaged structure factor
\begin{equation}
S(\bm q) = \frac{1}{N\beta} \sum_{ij} e^{-i{\bm q}\cdot ({\bm r}_i-{\bm r}_j)}  \int_0^{\beta} d\tau~  \langle S_i^z(\tau)S_j^z(0)\rangle,
\end{equation}
with $N = n\times L$ corresponding to the total number of sites. $[...]_{\rm av}$ denotes the disorder average, which is performed over 300-600 disorder realizations.

\section{Correlation length of site-diluted ladders}

\subsection{Effective Interactions among Localized Moments}
\label{s.LM}

Antiferromagnetic even-leg ladders without doping display a rung-singlet ground state \cite{DagottoR96} with a finite gap $\Delta$ to triplet excitations. In this state, the $n$ spins on the same rung preferentially form a singlet state, and therefore effectively decouple from the rest of the ladder. This leads to exponentially decaying correlations in the direction of the legs, characterized by a finite correlation length $\xi_0$.

At low enough concentration of dopants, the main effect of site dilution on an even-leg ladder is that of turning the number of spins on a rung from even to odd: in this situation, the state on the rung turns from a singlet into a \emph{doublet}, which corresponds to an effective $S=1/2$ localized moment (LM). This $S=1/2$ moment remains exponentially localized (over a characteristic length $\xi_0$) close to the impurity site \cite{Sandviketal97, Mikeskaetal97}, but its finite overlap with other LMs leads to an effective interaction which generically decreases exponentially with the distance. In the case of dominant rung interactions, $J_r \gg J_l$, the interaction between LMs is appropriately described within second-order perturbation theory (in $J_l/J_r$) as resulting from the exchange of virtual massive triplets between two LMs \cite{SigristF96}. In the case of a 2-leg ladder, in which we can identify the location of a LM with a spin site next to a missing rung partner (dangling spin), the effective coupling between LMs has the form of a SU(2)-invariant Heisenberg interaction, and at large separation between LMs, $|{\bm r}_i - {\bm r}_j| \gg 1$, the coupling strength takes the form \cite{SigristF96}
\begin{equation}
J^{({\rm eff})}_{ij} \sim  (-1)^{i+j+1}~ \frac{J_l^2}{\Delta}~ \frac{\exp(-|{\bm r}_i - {\bm r}_j|/\xi_0)}{\sqrt{|{\bm r}_i - {\bm r}_j|/\xi_0}},
\label{e.Jeff}
\end{equation}
where the staggering factor takes value $-1$ if the two LMs belong to the same sublattice and $+1$ otherwise.

Eq.~\eqref{e.Jeff} holds strictly speaking only in the case in which $J_l \ll \Delta$, but a similar exponential decay of the effective LM coupling (\emph{without} the square-root denominator) has been observed numerically in the case of much stronger leg interaction \cite{Mikeskaetal97}, as \emph{e.g.} for $J_l = J_r$, for which $\Delta \approx 0.53 J_l$. Moreover the above expression is numerically found to account for the decay of the effective couplings already for moderate distances, $|\bm r_i - \bm r_j| \gtrsim \xi_0$. Hence, towards an effective model of the site-diluted ladders, we will assume in the following that the exponential decay of LM couplings with decay rate $\xi_0$ remains valid even when perturbation theory is no longer applicable -- which is the case of interest. In this paper, we deliberately choose $J_r \leq J_l$ to have a sizable correlation length $\xi_0$, for reasons which will become clear in the following. Our assumption then implies that non-perturbative effects only affect the prefactor to the exponential in Eq.~\eqref{e.Jeff}. In addition, we will assume that the exponential decay present in Eq.~\eqref{e.Jeff} sets in for distances between LMs of the order of $\xi_0$.

In the case of $n$-leg ladders with $n>2$ a detailed theory of effective LM couplings is not available to our knowledge, but for widely spaced vacancies, the extended structure of the localized doublet becomes irrelevant, and for $J_l\ll J_r$ a perturbation approach analogous to that of 2-leg ladders should be applicable, leading to effective staggering couplings between LMs with exponential dependence on the distance.

\subsection{Statistical properties of the couplings for 2-leg ladders}
\label{s.PJeff}

\begin{figure}
\includegraphics[width=9.5cm,angle=0]{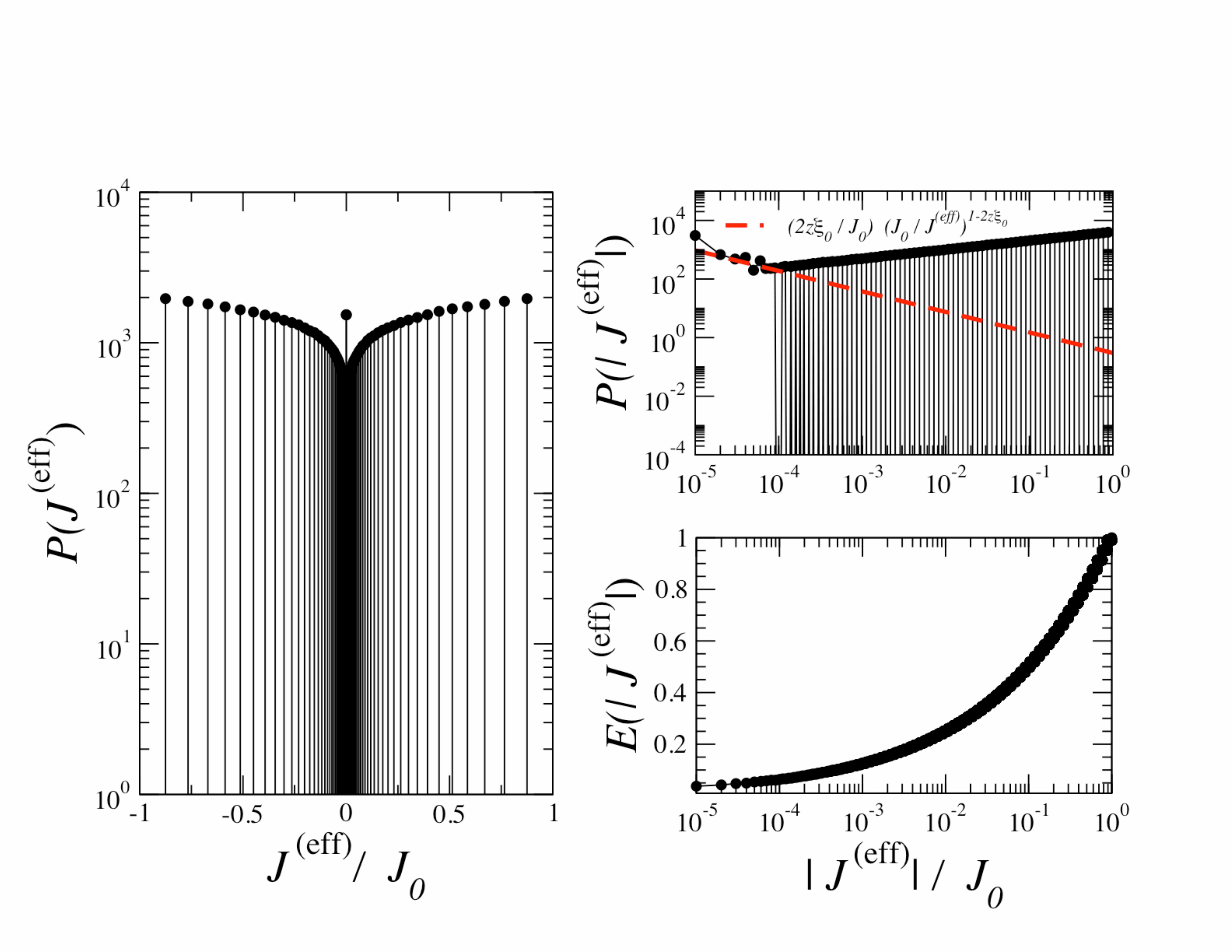}
\includegraphics[width=9.5cm,angle=0]{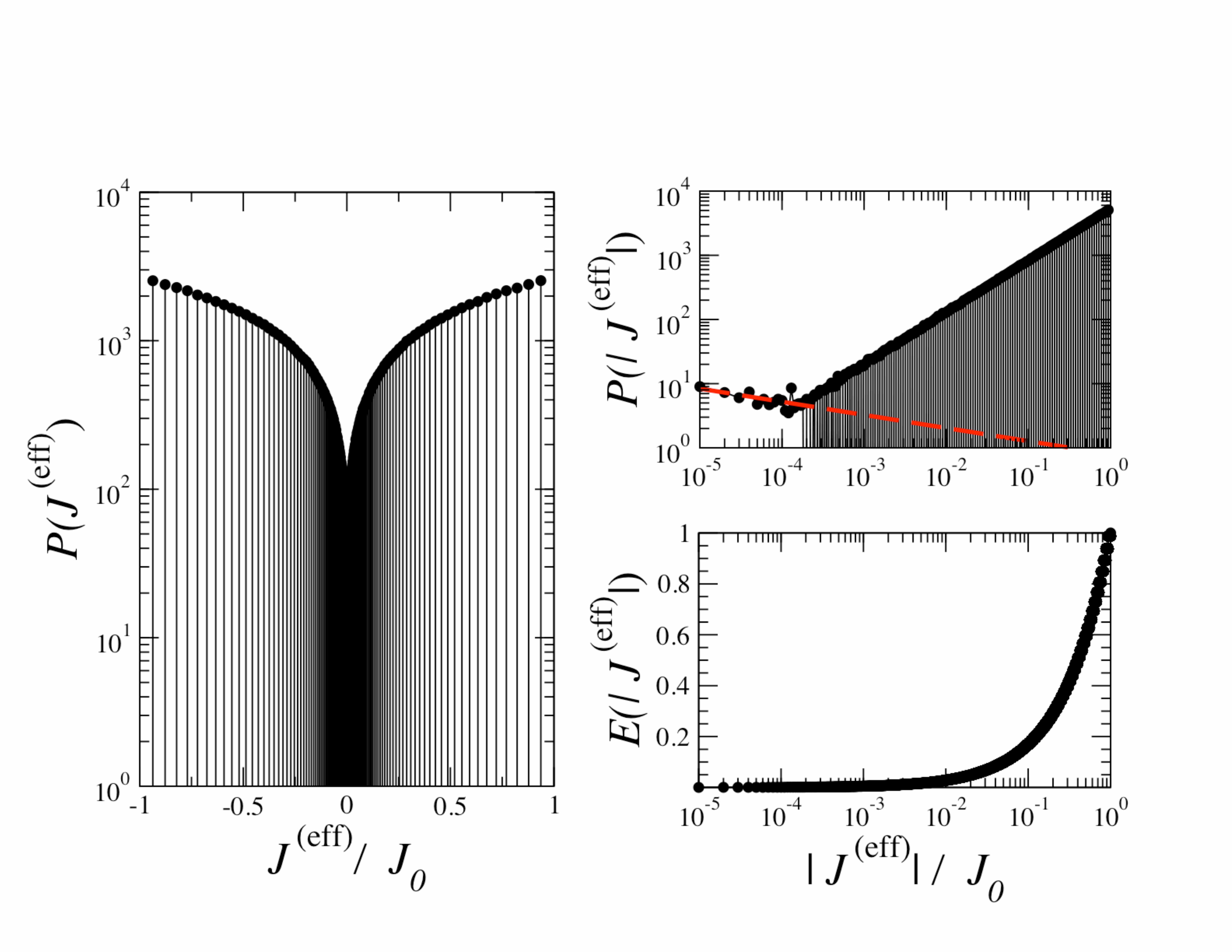}
\caption{\label{f.PJeff-numerical} (color online). Probability distribution ($P$) and cumulative distribution ($E$) of the effective couplings $J^{\rm (eff)}$ between nearest-neighbor localized moments in site-diluted 2-leg ladders, with parameters $z=2\%$, $\xi_0 = 7.5$ (upper panels), and $z=2.5\%$, $\xi_0 = 15.3$ (lower panels). The red dashed lines show the probability distributions in the continuum approximation according to Eq.~\eqref{e.PJeff}.}
\end{figure}

In the case of low doping, $z \ll1 $, the probability distribution of having two nearest-neighboring LMs at a distance $d$ along the leg direction in a 2-leg ladder is given by
\begin{equation}
P(d) = 2z~\exp(-2zd)
\end{equation}
with a corresponding average $\langle d \rangle = 1/(2z)$. As found numerically in Ref.~\onlinecite{Mikeskaetal97}, the effective coupling has the form $J^{({\rm eff})}(d) \approx J_0 \exp(-d/\xi_0)$ for $d\gtrsim \xi_0$. As an approximation, one may assume that the same exponential behavior survives for $d\lesssim \xi_0$. Then the probability distribution for the strength of the couplings between nearest neighboring LMs reads
\begin{eqnarray}\label{e.PdPJsum}
P\left(J^{({\rm eff})}_{\rm nn}\right) &=& \sum_{d=0}^{\infty} ~P(d) ~
\delta[J^{({\rm eff})}_{\rm nn} - J_0 \exp(-d/\xi_0)].
\end{eqnarray}
This result suggests the possibility of extracting analytically the probability distribution for $J^{({\rm eff})}_{\rm nn}$, as done in Ref.~\onlinecite{SigristF96}. To this end, one may further take the continuum approximation, namely, approximate the summation in Eq.~\eqref{e.PdPJsum} by an integral:
\begin{eqnarray}
P\left(J^{({\rm eff})}_{\rm nn}\right) &\approx& \int dl~ P(l) ~
\delta[J^{({\rm eff})}_{\rm nn} - J_0 \exp(-l/\xi_0)]~.
\label{e.PdPJ}
\end{eqnarray}
This allows then to write
\begin{equation}
 P\left(J^{({\rm eff})}_{\rm nn}\right) \approx \frac{1-\gamma}{2J_0} \left(\frac{J_0}{J^{({\rm eff})}_{\rm nn}}\right)^{\gamma},
\label{e.PJeff}
\end{equation}
where $\gamma = 1- 2z \xi_0$. Hence, within the continuum approximation the n.n. couplings obey a simple power-law distribution.\cite{SigristF96} The average absolute value of the coupling between neighboring LMs takes then the form \cite{SigristF96}:
\begin{equation}
[ |J^{({\rm eff})}_{\rm nn} |]_{\rm av} = \frac{1-\gamma}{2-\gamma} ~J_0 = \frac{2z\xi_0}{1+2z\xi_0} ~J_0.
\label{e.Jav}
\end{equation}

However, a critical analysis of the above derivation shows that the continuum approximation is problematic. In fact, it requires that the distances $d$ giving a significant contribution to the sum in Eq.~\eqref{e.PdPJsum} be $d\gg 1$, which is in contradiction with the fact that $P(d)$ decreases exponentially with $d$; the characteristic decay length is $\langle d\rangle = (2z)^{-1}$, which for $z \sim 2\%$ takes values $\sim 20$.

Fig.~\ref{f.PJeff-numerical} shows the distribution $P(J^{({\rm eff})}_{\rm nn})$ determined numerically according to Eq.~\eqref{e.PdPJsum}, {\it i.e.}, by sampling the \emph{discrete} distribution lengths $P(d)$, for $z = 2\%$, $\xi_0 = 7.5$ (corresponding to $J_r/J_l= 1/2$, and giving $\gamma = 0.7$), and $z = 2.5\%$, $\xi_0 = 15.3$ (corresponding to $J_r/J_l= 1/4$, and giving $\gamma = 0.235$) -- these parameter sets will be relevant for our study of correlations in the following. We notice that the distribution shown in Fig.~\ref{f.PJeff-numerical} is only quantitatively correct when $d \gtrsim \xi_0$, namely for $J_{\rm eff}/J_0 \lesssim 1/e$, while it is only an approximation otherwise.

It is clear that for $d \gtrsim \xi_0$, the distribution of n.n. couplings obtained numerically deviates strongly from the prediction of  Eq.~\eqref{e.PJeff}. Over most of its support, the distribution of n.n. couplings has a fundamentally \emph{discrete} structure, due to the fact that the largest values of $J^{({\rm eff})}_{\rm nn}$, which also take the largest probabilities, are associated with short separations $d$ among n.n. LMs. These probabilities \emph{grow} as a power law of the absolute value of the coupling, opposed to what predicted by Eq.~\eqref{e.PJeff} for $\gamma > 0$. Obviously a coarse graining on the $\delta$-peaks of the exact distribution would gradually reconstruct the shape of Eq.~\eqref{e.PJeff}, but given the finite separation between the $\delta$-peaks even in the thermodynamic limit, this coarse graining is not justified over a large range of $J^{({\rm eff})}_{\rm nn}$. In fact the $\delta$-peaks become dense only in the limit  $J^{({\rm eff})}_{\rm nn} \to 0$, which corresponds to larger and larger separations among the LMs, and hence to lesser and lesser probabilities. Only in that limit the continuum approximation appears legitimate (in fact numerically we recover the analytical prediction for $J^{({\rm eff})}_{\rm nn} \lesssim 10^{-4} J_0$ due to the finite width of our bins, $J^{({\rm eff})}_{\rm nn} = 10^{-5} J_0$).

In conclusion, for  $ z \gtrsim 1\%$ and $\xi_0 \sim 10$, $P(J^{({\rm eff})}_{\rm nn})$ has a remarkable structure: it assigns the largest probability to the range in which $J^{({\rm eff})}_{\rm nn}$ is a \emph{discrete} variable, preventing a straightforward approximation of the distribution with a continuous function. This will have significant consequences on the analysis of doped ladder systems in relation with models of randomly coupled spins.

\begin{figure}
\includegraphics[width=7.5cm,angle=0]{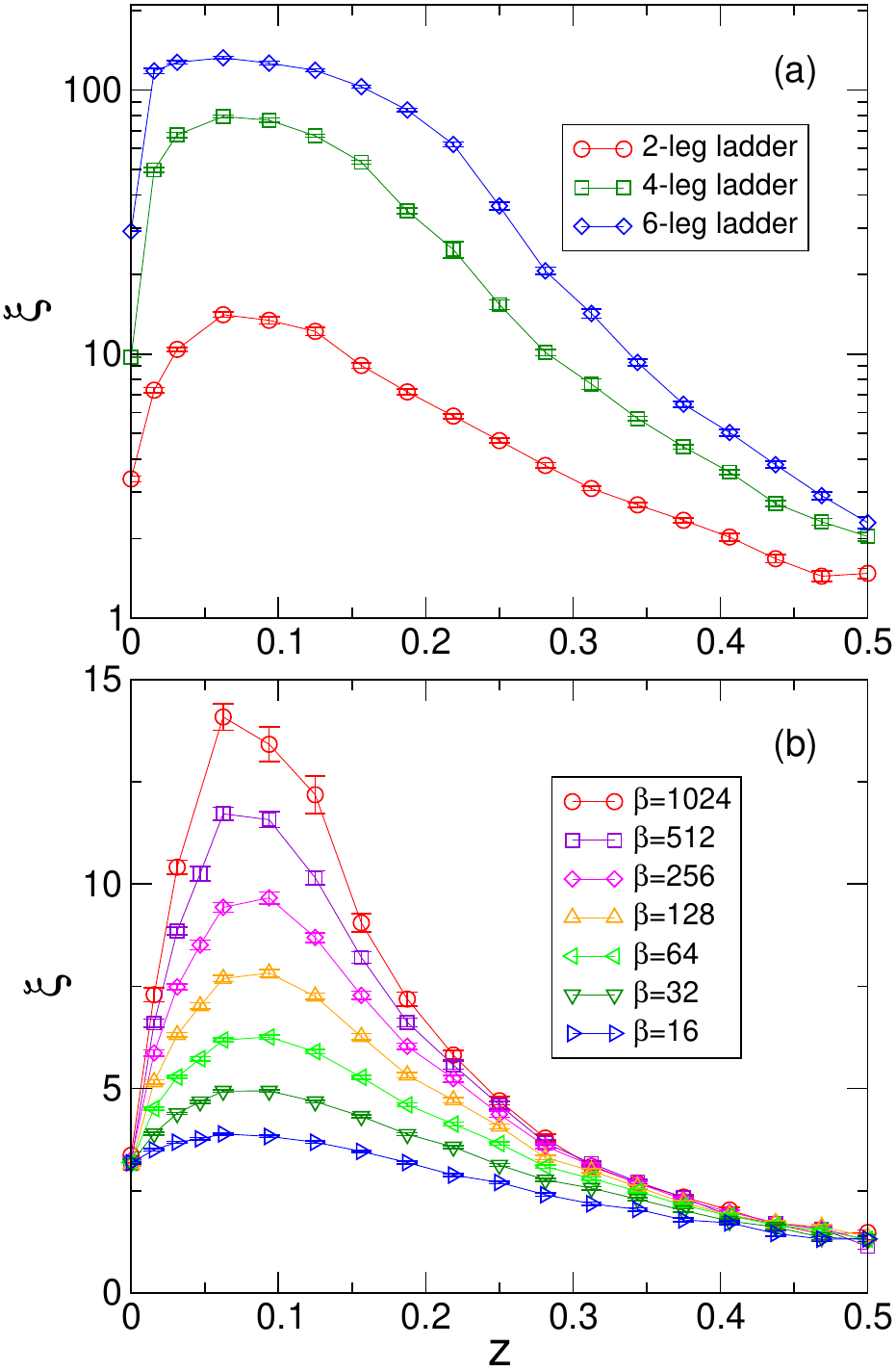}
\caption{\label{f.sitedil} (color online). Correlation length $ \xi $ of site-diluted ladders with $L=128$ and $J_l = J_r = J$. Upper panel: comparison between 2-leg, 4-leg and 6-leg ladders at inverse temperature $\beta J = 1024$. Lower panel: correlation length of the 2-leg ladder at different temperatures.}
\end{figure}

\subsection{Correlations as a function of doping}
\label{s.corrdoping}

In both cases of $n=2$ and $n>2$ the unfrustrated effective couplings between LMs can lead to the formation of a \emph{gapless} ground state with algebraically decaying correlations between the sites, and hence an infinite correlation length. When looking at a fixed finite temperature $T$ the correlation length is necessarily finite for one-dimensional SU(2) invariant systems with short-range interactions, but it is expected to increase because the average coupling between LMs increases with $z$, as in Eq.~\eqref{e.Jav}, so that the effective temperature $T/[ |J^{({\rm eff})}_{\rm nn}| ]_{\rm av}$ decreases.

Despite the increase of correlations due to disorder, the range of correlations is fatally bounded by the fact that any finite amount of site dilution will break an infinite $n$-leg ladders into finite segments, of characteristic length $\langle l \rangle \approx z^{-n}$ (corresponding to the inverse linear density of rungs which are fully removed by doping). When considering doped ladders at finite temperatures no geometric bound on the correlation length (neither coming from finite-size effects nor from ladder fragmentation) is present as long as the correlation length satisfies the condition
\begin{equation}
\xi(T) \ll L \ll \langle l \rangle .
\label{e.condition}
\end{equation}
It is clear that such a condition can only be satisfied when the doping is sufficiently weak. Hence, for any fixed temperature, upon increasing the doping concentration the correlation length $\xi(T;z)$ will cross over from a growing behavior at low doping $z\ll 1$ to a decreasing behavior for $z \gtrsim 0.1$, hence going through a maximum for a given optimal doping value $z^* = z^*(J_{r}/J_l;n)$, which is quite stable at sufficiently low temperatures. This is indeed seen in Fig.~\ref{f.sitedil}(a), in which we observe \cite{footnote1} that, for $J_{r}=J_l$, $z^* \sim 0.06$ for $n=2$, 4, and 6.

For any finite value of doping and at arbitrarily low temperatures, the correlation length is upper bounded by the average length of the ladder segments, $\langle l \rangle$. Yet, as shown in Fig.~\ref{f.sitedil}(b), at low enough doping the upper bound is reached extremely slowly in temperature - in fact, for the 2-leg ladder at $T = J_l/1024$ and $J_l = J_r$, the correlation length is still not saturated to its finite $T=0$ value for $z \lesssim 0.25$. It is easy to understand that this occurs because the average effective coupling $[ |J^{({\rm eff})}_{\rm nn}| ]_{\rm av}$ decreases as $z$ decreases, leading to exceedingly weak correlations even at extremely low temperatures.

This means that for sufficiently low doping, the condition  $\xi(T) \ll  \langle l \rangle$ is actually satisfied over all numerically accessible low temperatures (and quite possibly also over all experimentally accessible temperatures), so that the segmented nature of the doped ladders become in fact \emph{irrelevant}. In the following we will focus our discussion on this regime.

\subsection{Doped 2-leg ladders and the random-exchange Heisenberg model}

As discussed in section \ref{s.PJeff}, from the statistical point of view the system of interacting LMs randomly distributed on the bipartite 2-leg ladder can be approximated by a system of $S=1/2$ spins interacting with randomly distributed couplings, following an intrinsically \emph{discrete} distribution. In general terms, a doped 2-leg ladder represents a physical realization of the so-called random-exchange Heisenberg model (REHM) \cite{Westerbergetal95, Westerbergetal97} , but with a special structure of the coupling distribution.

\begin{figure}
\includegraphics[width=7.5cm,angle=0]{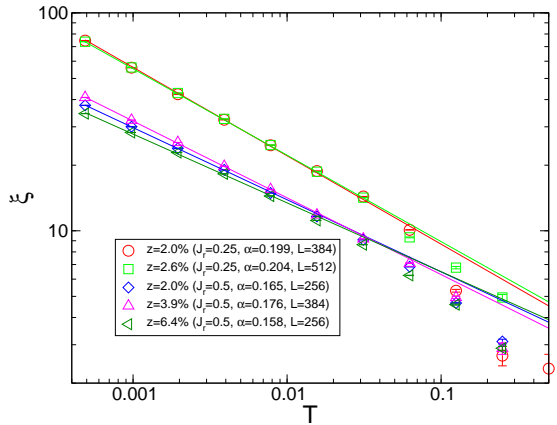}
\caption{\label{f.xiT} (color online). Temperature dependence of the correlation length of site-diluted 2-leg ladders with $J_l=1$, and power-law fits at low temperature.}
\end{figure}

Numerical real-space renormalization group studies \cite{Westerbergetal95, Westerbergetal97} show that the REHM {color{red} for a \emph{continuous} distribution of couplings $P(\tilde J)$} has two regimes:
\begin{itemize}
\item \emph{Universal regime}: if the distribution is not singular, or has a power-law singularity $P(\tilde J) \sim J^{-\gamma}$ with $\gamma < \gamma_c$ (where $0.65 \lesssim \gamma_c \lesssim 0.75$), the low-temperature thermodynamics of the REHM is dominated by a unique fixed point; the correlation length diverges as $T \to 0$ with a universal scaling exponent, $\xi \sim T^{-2\alpha}$ with $\alpha = 0.22\pm 0.01$. This result has been confirmed by quantum Monte Carlo \cite{Frischmuthetal99} for a non-singular distribution (square box, $\gamma = 0$).
\item \emph{Non-universal regime}: for a more strongly singular distribution $\gamma > \gamma_c$ the renormalization group flows to a non-universal fixed point, which depends on the initial coupling distribution. Studies on  doped coupled spin-Peierls chains using Real Space Renormalization Group and SSE QMC in this regime also confirm that the exponent $\alpha$ depends on the doping concentration\cite{Laflorencie05}.
\end{itemize}

As discussed in Sec.~\ref{s.PJeff}, the distribution of couplings between LMs in a doped 2-leg ladder is intrinsically discrete, approximating a continuous distribution only for vanishing couplings. It is therefore interesting to ask whether the doped 2-leg ladder reproduces the known physics of the REHM model with a continuous distribution of couplings. If a continuous approximation to the coupling distribution is in order, as in Eq.~\eqref{e.PJeff}, the system should exhibit a \emph{universal REHM behavior} for $2z \xi_0 > 1- \gamma_c$, and a non-universal behavior for $2z \xi_0 < 1- \gamma_c$. In other words, ladders with sufficiently strong doping $z$ or sufficiently large correlation length $\xi_0$ in the undoped limit should exhibit universal scaling properties at low temperature. To our knowledge the only numerical study of correlations in site-diluted ladders is Ref.~\onlinecite{GrevenB98}, which, making use of quantum Monte Carlo, investigates ladders with $J_r/J_l = 1/2$ and dopings $z = 0.01$, 0.04 and 0.1. Given that $\xi_0 \approx 7.5$ in this case \cite{GrevenB98}, for $z = 0.04$ and 0.1 one obtains $1 - 2z \xi_0 = 0.4 $ and $-0.5$ respectively, which appear to be both safely on the universal side of the REHM with continuous couplings. Indeed a fit to the low-temperature behavior of the correlation length to the scaling law $\xi = A(z) + B(z)~ T^{-2\alpha(z)}$ gives an estimate of $\alpha = 0.2 \pm 0.025$ (independent of $z$) which appears consistent with the predictions of Refs.~\onlinecite{Westerbergetal95, Westerbergetal97}.

\subsection{Low-temperature scaling of the correlation length}

\begin{figure}%{h}
\includegraphics[width=5.8cm,angle=-90]{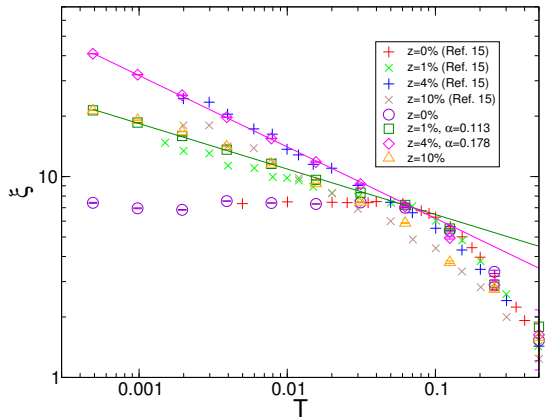}
\caption{\label{f.greven} (color online). Comparison of our data with those by Ref.~\onlinecite{GrevenB98}. Our data refer to ladders with $L=384$, and $J_l=2J_r=1$.}
\end{figure}

We have performed QMC simulations to extensively study the low-temperature behavior of the correlations in site-diluted 2-leg ladders. Our QMC results, however, do not support the conclusion in Ref.~\onlinecite{GrevenB98} about a universal REHM behavior of diluted 2-leg ladders. We investigate doped ladders at similar concentrations with respect to those of Ref.~\onlinecite{GrevenB98}, down to significantly lower temperatures - Ref.~\onlinecite{GrevenB98} stops at $T = J_l/500$, while we descend down to  $T = J_l/2048$. A collection of our results is shown in Fig.~\ref{f.xiT}. In qualitative agreement with what expected from the theory of the REHM model, we also observe the clear onset of a low-temperature power law scaling of the type $T^{-2\alpha}$. Yet our central observation is that this power-law scaling exhibits a \emph{non-universal, doping dependent} exponent $\alpha=\alpha(z)$. For all the parameter sets we considered, a fit of the low-temperature correlation length provides values of $\alpha$ which are systematically below the one exhibited by the universal regime of the REHM with continuous couplings.

The data shown in Fig.~\ref{f.xiT} refer to 2-leg ladders with $J_r/J_l = 1/4$ and $z=2.0 \%$ and $2.6 \%$, with corresponding $\gamma \approx 0.39$  and $0.235$ respectively (here we use $\xi_0 = 15.3$, which we estimate independently with QMC); and 2-leg ladders with  $J_r/J_l = 1/2$ (as in Ref.~\onlinecite{GrevenB98}) and $z = 2.0\%$, $3.9\%$, and $6.4\%$, corresponding to $\gamma \approx 0.7$, $0.415$, and $0.04$ respectively. Although all the values of $\gamma$ fall below the critical value $\gamma_c$, as shown in Fig.~\ref{f.xiT}, in fact, all the corresponding data sets for the correlation length display a \emph{non-universal} power-law scaling. In particular, we observe that data sets with close values of the parameter $\gamma$ (\emph{e.g.} $J_r/J_l = 1/4$, $z=2.0 \%$ with $\gamma \approx 0 .39$,  and $J_r/J_l = 1/2$, $z = 3.9\%$, with $\gamma \approx 0.415$) show a clearly different scaling. We also critically revisit the parameter sets explored in Ref.~\onlinecite{GrevenB98}. A comparison to our data, as done in Fig.~\ref{f.greven}, shows that in the data of Ref.~\onlinecite{GrevenB98} the significant scattering due to the statistical uncertainty is masking the correct asymptotic scaling regime at low temperature.

\begin{figure}
\includegraphics[width=7.5cm,angle=0]{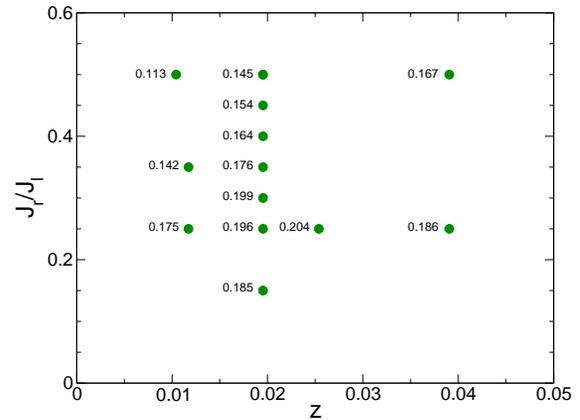}
\caption{\label{ph.diagram} (color online). Values of the $\alpha$ exponent extracted from the low-temperature behavior of site-diluted 2-leg ladders.}
\end{figure}

A comprehensive summary of our results on the low-temperature scaling of the correlation length is provided in  Fig.~\ref{ph.diagram}, where we report the values of the $\alpha$ exponent obtained for different doping values and $J_r/J_l$ ratios. The values of doping concentrations and ratios were chosen such that the average distance between LMs $\langle d \rangle$ is at least larger than $\xi_0$. The LMs, therefore, are not overlapping on average and can be  described by an effective REHM with couplings obeying Eq.~\eqref{e.Jeff}. Furthermore, for all the points shown in Fig.~\ref{ph.diagram}, $\gamma < \gamma_c$, we do not observe a universal $\alpha$ value. Indeed, it is found that $\alpha$ depends not only on the doping concentration but also on the ratio of couplings $J_r/J_l$. And an $\alpha$ exponent close to that of the universal REHM regime is only observed in a narrow region around  $J_r/J_l \sim 1/4$ and $z \sim 2\%$.

\subsection{Discussion}

The above results lead us to conclude that doped 2-leg ladders do not realize in general the universal physics of the REHM with a continuous coupling distribution. In fact, the coupling distribution for the LMs in doped 2-leg ladders has a fundamentally discrete structure, approximating a continuous one only for vanishing couplings. The weight assigned by the distribution to strong discrete couplings is dominant over that of weak, quasi-continuous couplings in the case of sizable doping $z$ (suppressing the probability of large separation between the LMs) and of large correlation length $\xi_0 \gg 1$ (increasing the strength of the couplings between LMs, the distance being fixed). Nonetheless, this is the regime in which the parameter $\gamma = 1 - 2z\xi_0$ is small, and in which one would expect the universal REHM physics to be manifested in the continuum approximation. On the other hand, the  quasi-continuous part of the distribution (corresponding to very small couplings) acquires a bigger weight for small $z$ and $\xi_0$, which determine a small average coupling according to Eq.~\eqref{e.Jav} and hence increase the probability of the small values of the coupling. In this regime $\gamma $ is closer to 1, which leads to too strong a singularity in the probability distribution for the universal regime of the REHM with continuous couplings to be manifested.

As a consequence, we generally observe a low-temperature power-law scaling of the correlation length with doping-dependent exponents. As far as the numerical results are concerned, this scaling appears to be the asymptotic one for $T\to 0$, given that it sets in for temperatures well below the average LM coupling. On the other hand, according to a real-space renormalization group (RSRG) approach \cite{Westerbergetal95}, the system at a lower temperature is increasingly sensitive to the part of the distribution related to the weaker couplings; given that this part is the one which best approximates a continuous distribution, one could naturally expect that at very low temperatures the physics of the REHM with continuous couplings be reproduced by the doped 2-leg ladders. This argument would then suggest that, when lowering the temperature, the system will sooner or later attain the universal regime of the REHM with continuous couplings if $\gamma < \gamma_c$. Nonetheless one can argue that in a doped 2-leg ladder the RSRG flow has to be necessarily stopped at a finite length, corresponding roughly to the average segment length $\langle l \rangle$. In practice this imposes an upper bound to the correlation length and a lower bound to the temperature through the condition Eq.~\eqref{e.condition}; these bounds might prevent one from attaining the low-temperature regime at which universal REHM physics would manifest itself.

Finally, our study leaves one open question concerning the role of discreteness of the initial coupling distribution for the physics of the REHM. The numerical RG study of Ref.~\onlinecite{Westerbergetal97} only addresses initial continuous distributions; a systematic RG study of the flow starting from realistic distributions stemming from the doped-ladder physics would be highly desirable.

\section{Bond-diluted ladders}

In the following, we consider the doping- and temperature dependence of correlations in bond-diluted ladders, which show a marked difference with respect to site dilution. This is due to a fundamental geometric aspect which distinguishes site and bond dilution: diluting a site leaves a single unpaired spin, giving rise to a LM, while eliminating a bond leaves always \emph{two} unpaired spins, located on different sublattices. The corresponding LMs are therefore interacting with an effective antiferromagnetic coupling, mediated by the the shortest path of bonds connecting them. In the following, we will focus for simplicity on the case $J_l = J_r = J$.

\subsection{2-leg ladders}

\subsubsection{Evolution of correlations with doping}

\begin{figure}
\includegraphics[width=7.5cm,angle=0]{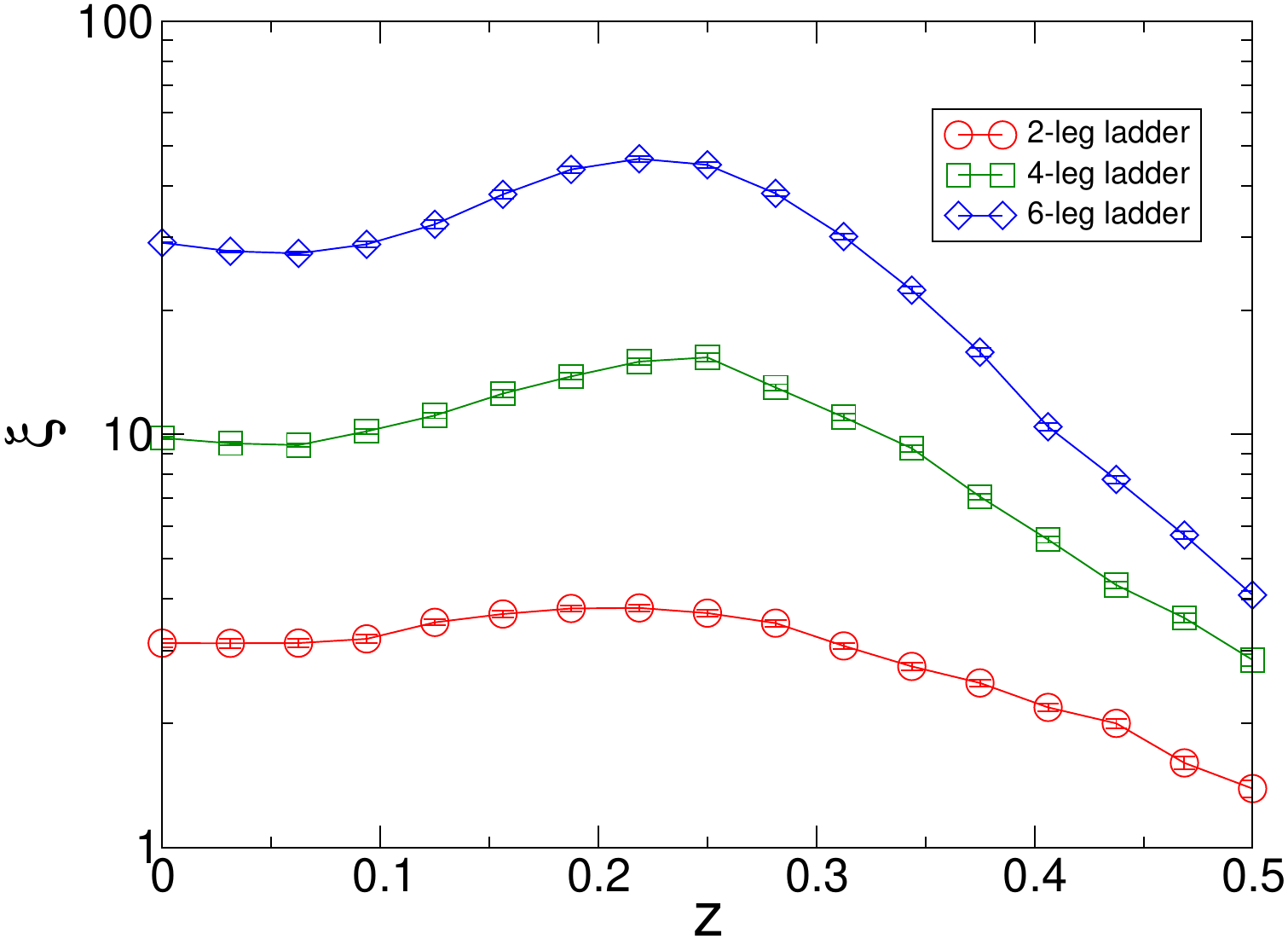}
\caption{\label{f.bonddil-fixT} (color online). Correlation length of bond-diluted 2-, 4-, and 6-leg ladders with $L = 128$, $J_l = J_r = J$, and temperature $T= J/1024$}
\end{figure}

Fig.~\ref{f.bonddil-fixT} shows the evolution of the correlation length at a fixed, low temperature $T = J/1024$. A striking difference with respect to the case of site dilution, shown in Fig.~\ref{f.bonddil-fixT}, is that the correlation length does not increase immediately with doping: it remains essentially constant for $n=2$, while it even \emph{decreases} for $n>2$.

\begin{figure}
\includegraphics[width=8cm,angle=0]{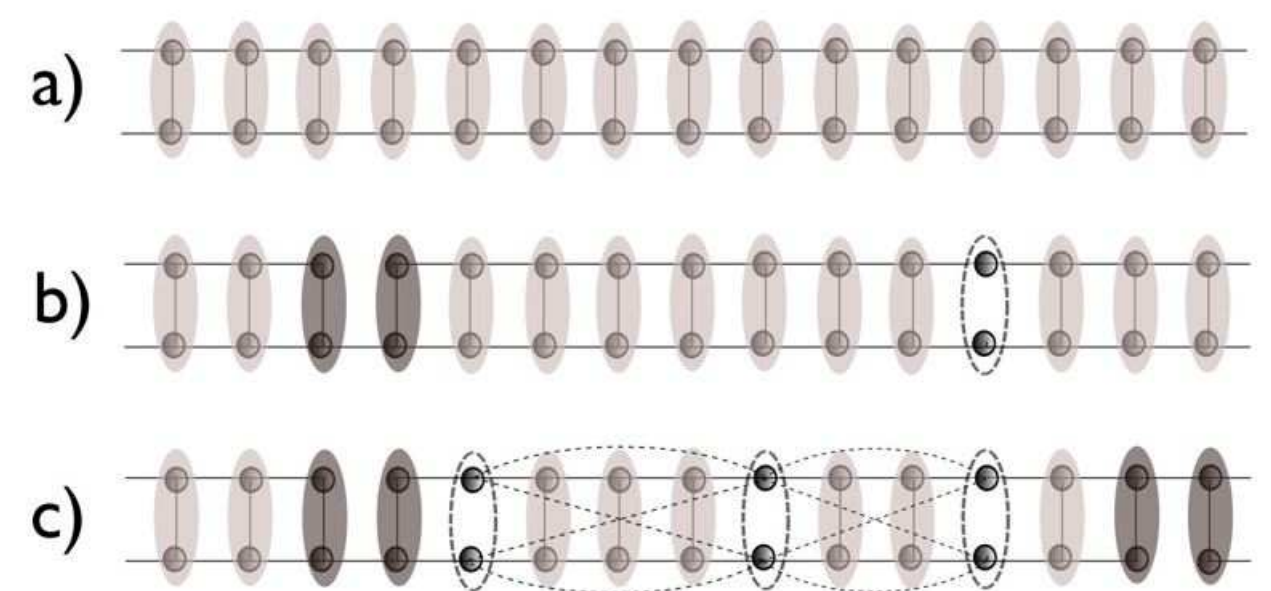}
\caption{\label{f.bonddil-cartoon} (color online). Sketch of the bond dilution effects on 2-leg ladders. a) pure ladder; b) ladder at low dilution: enhancement of rung singlets by leg-bond dilution, and low-energy singlets formed between LMs after rung-bond dilution; c) ladder at stronger dilution: couplings between LMs belonging to different diluted rungs.}
\end{figure}

In the case of a 2-leg ladder, we can understand qualitatively the behavior of correlations by considering that bond dilution has two fundamentally different effects (sketched in Fig.~\ref{f.bonddil-cartoon}(b)):
\begin{itemize}
\item if a rung bond is diluted, two LMs are liberated but still interact antiferromagnetically, so that they can screen each other to form a rung singlet at a lower energy $J^{\rm(eff)}_{_{\rm rung}}$ (corresponding to the effective coupling between LMs which are rung neighbors). This screening is very effective at low dilution because in that case the  $J^{\rm(eff)}_{_{\rm rung}}$ interaction is largely dominant over the interaction between two LMs belonging to different rungs, due to the large spacing of two diluted rung bonds -- this spacing is $3/z$ (given that only one bond in three is a rung bond);
\item if a leg bond is diluted, the rung-singlet state on the two adjacent rungs is further reinforced because its connectivity to the other rungs is lowered.
\end{itemize}

The dilution of rung bonds has therefore the effect of lowering locally the gap over the ground state, whereas the dilution of leg bonds has the effect of increasing it \cite{noteonlocalgap}. The fact that there are twice as many leg bonds as rung bonds suggests that the second effect dominates over the first. Therefore the spin gap of the undoped ladder is locally preserved or even enhanced, leading to short-range correlations even in presence of bond dilution. Nonetheless, standard consideration on rare-event physics lead to conclude that even a very weak bond dilution leads the system to a \emph{gapless, Griffiths-like phase}. In fact, the gap can close locally if rare regions appear in which only rung bonds are diluted -- the limiting case being that of the formation of local strands made of two uncoupled $S=1/2$ chains. These exponentially rare regions lead to the closing of the spin gap in the thermodynamic limit; yet global correlations remain short ranged, due to the localized nature of the rare, locally gapless regions. This Griffiths phase is reminiscent of what has been observed by some of us in the Heisenberg model on the  inhomogeneously bond-diluted square lattice \cite{Yuetal05,Yuetal06}. In that case, the predominance of correlation-suppressing dilution was guaranteed by the inhomogeneity of doping probabilities (favoring the appearance of ladder-like or dimer-like structures).

\begin{figure}
\includegraphics[width=7.5cm,angle=0]{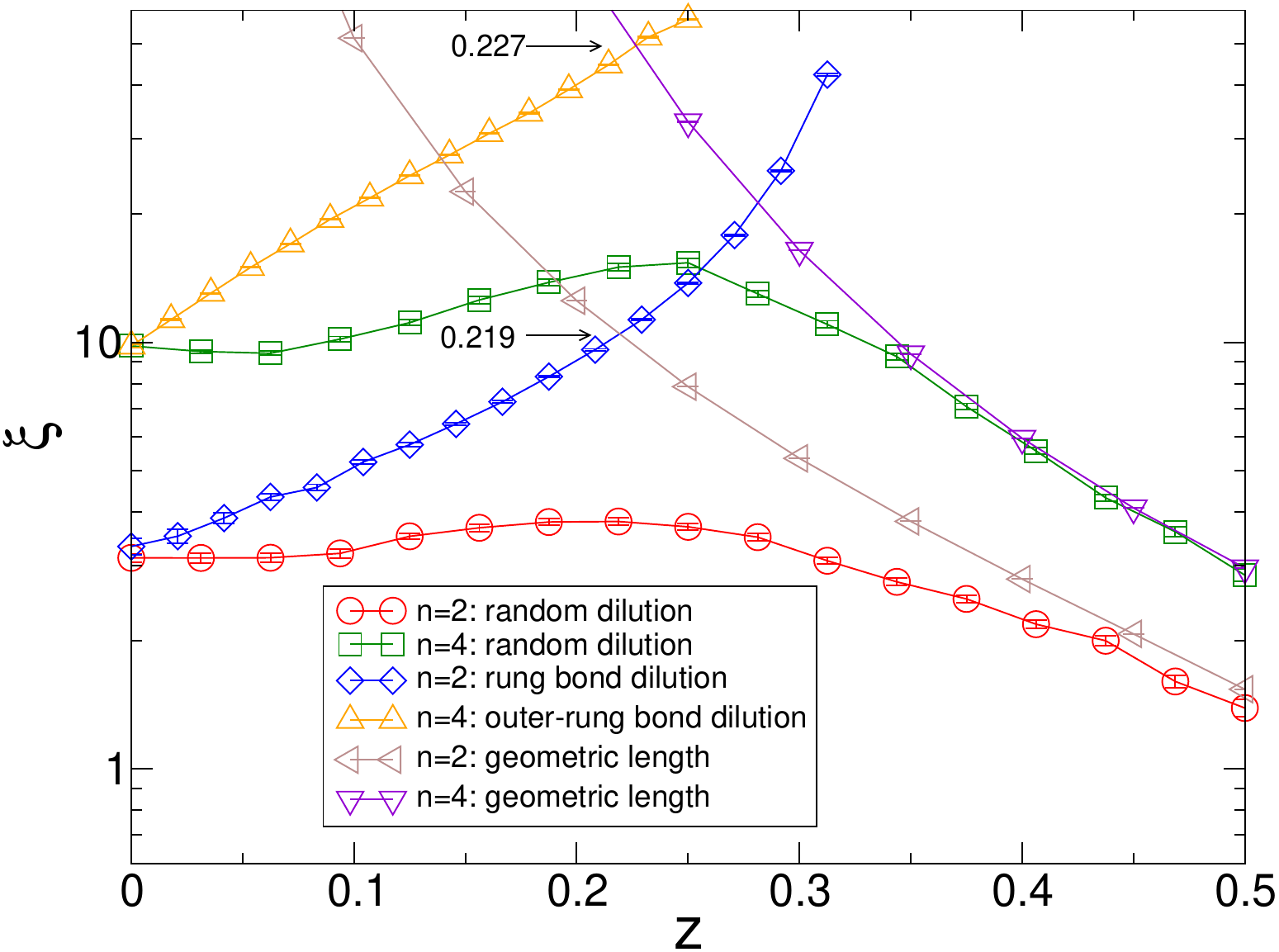}
\caption{\label{f.rungdilution} (color online). Correlation length for 2-leg and 4-leg ladders with dilution of (center-)rung bonds. The average length of ladders segments $\langle l \rangle$ is obtained upon homogeneous dilution. }
\end{figure}

For a sufficiently large bond dilution $z > z^*$ ($0.07 \lesssim z^* \lesssim 0.09$), the correlation length starts increasing with increasing $z$. This implies necessarily that the liberated LMs belonging to different rungs begin to correlate with each other (Fig.~\ref{f.bonddil-cartoon}(c), see also Refs. \onlinecite{Yuetal06, Yasuda06}). Such correlations appear because the spacing between diluted rung bonds decreases, and hence the couplings between LMs belonging to different rungs become of the same order of $J^{\rm(eff)}_{_{\rm rung}}$. We have isolated the effect of enhancement of correlations through rung-bond dilution by considering uniquely this form of dilution, namely by taking  $p^{(r)}_{i,m} = 0$ with probability $z$, while  $p^{(l)}_{i,m} = 1$ with probability 1. The resulting correlations $\xi_{\rm r}(z)$ (r = rung) as a function of $z$ for 2- and 4-leg ladders are shown in Fig.~\ref{f.rungdilution}, and they are seen to increase quite fast with dilution (following the approximate form $\xi_{\rm r}(z) \approx \xi_0 \exp(a z)$ for sufficiently low dilution). Yet in the original system the correlations induced by rung-bond dilution are upper-bounded by the characteristic length of the ladder segments $\langle l \rangle$, which we have estimated by generalizing an efficient algorithm recently developed for homogeneous percolation \cite{Stauffer85, Newman}. It is found that at finite $z$, $\langle l \rangle \approx a/z^2$. It follows the same asymptotic behavior as from the naive estimate $\langle l\rangle \sim 1/z^2$, but with a renormalized factor $a<1$. Hence the enhancement of correlations due to rung-bond dilution competes with this geometrical restriction on correlations for sufficiently strong doping. Quite remarkably, we observe that the optimal doping for the enhancement of correlations in the homogeneously doped 2-leg ladder coincides with the value of dilution ($z\approx 22\%$) at which the correlation length of the rung-only diluted ladder equals the average length of ladder segments $\langle l \rangle$. Therefore we can conclude that the non-monotonic behavior of the correlation length at intermediate dilution values is completely determined by the competition between the correlation enhancement through rung-bond dilution and the correlation suppression due to ladder fragmentation.

\begin{figure}
\includegraphics[width=8.5cm,angle=0]{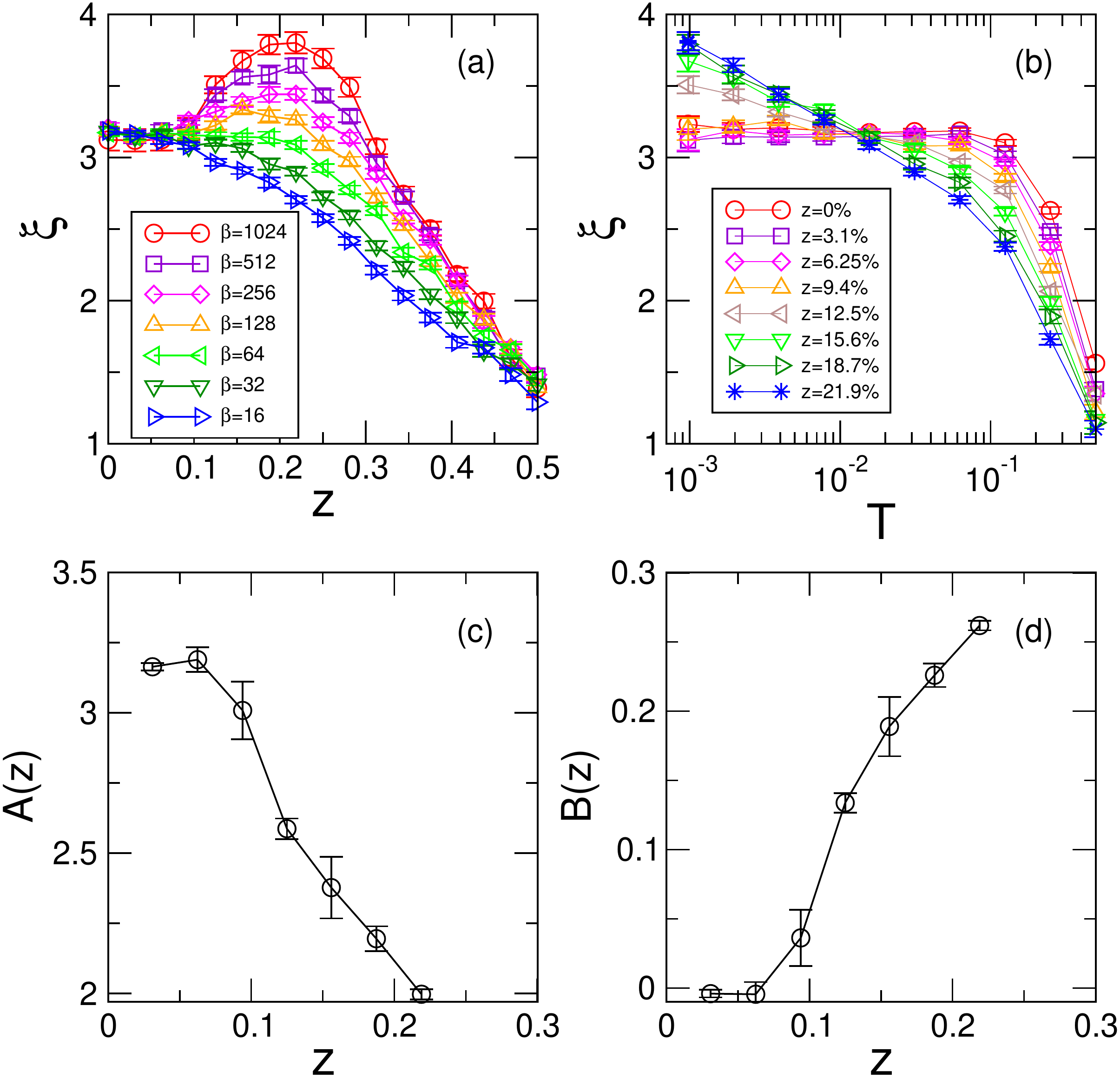}
\caption{\label{f.bonddilution.T} (color online). (a) Correlation length of bond-diluted 2-leg ladders with $L=128$ and $J_l=J_r=J$ at various temperatures. (b) Temperature dependence of the correlation length of above system. The curves with $z > 22\%$ are omitted due to $\xi$ is limited by the geometric length. (c) and (d) Fitting parameters of correlation length $\xi \approx A(z) + B(z)|\log(T/J)|$, as a function of bond dilution.}
\end{figure}

 \subsubsection{Evolution of correlations with temperature}

As discussed in the previous subsection, for $z = z^*$ correlations start to increase as a function of bond dilution. Fig.~\ref{f.bonddilution.T}(a) shows the doping dependence of the correlation length of 2-leg ladders for higher temperatures than those considered in Fig.~\ref{f.bonddil-fixT}: here we observe that the correlation length does not appear to evolve significantly with temperature for doping values $z \lesssim z^*$, whereas it becomes significantly dependent on temperature for  $z > z^*$. Moreover, the low-temperature dependence of the correlation length, see Fig.~\ref{f.bonddilution.T}(b), indicates that in this regime the correlation length grows logarithmically with decreasing temperature as
\begin{equation}
\xi \approx A(z) + B(z) |\log (T/J)|~.
\label{e.xi-bl}
\end{equation}
\begin{figure}
\includegraphics[width=7.5cm,angle=0]{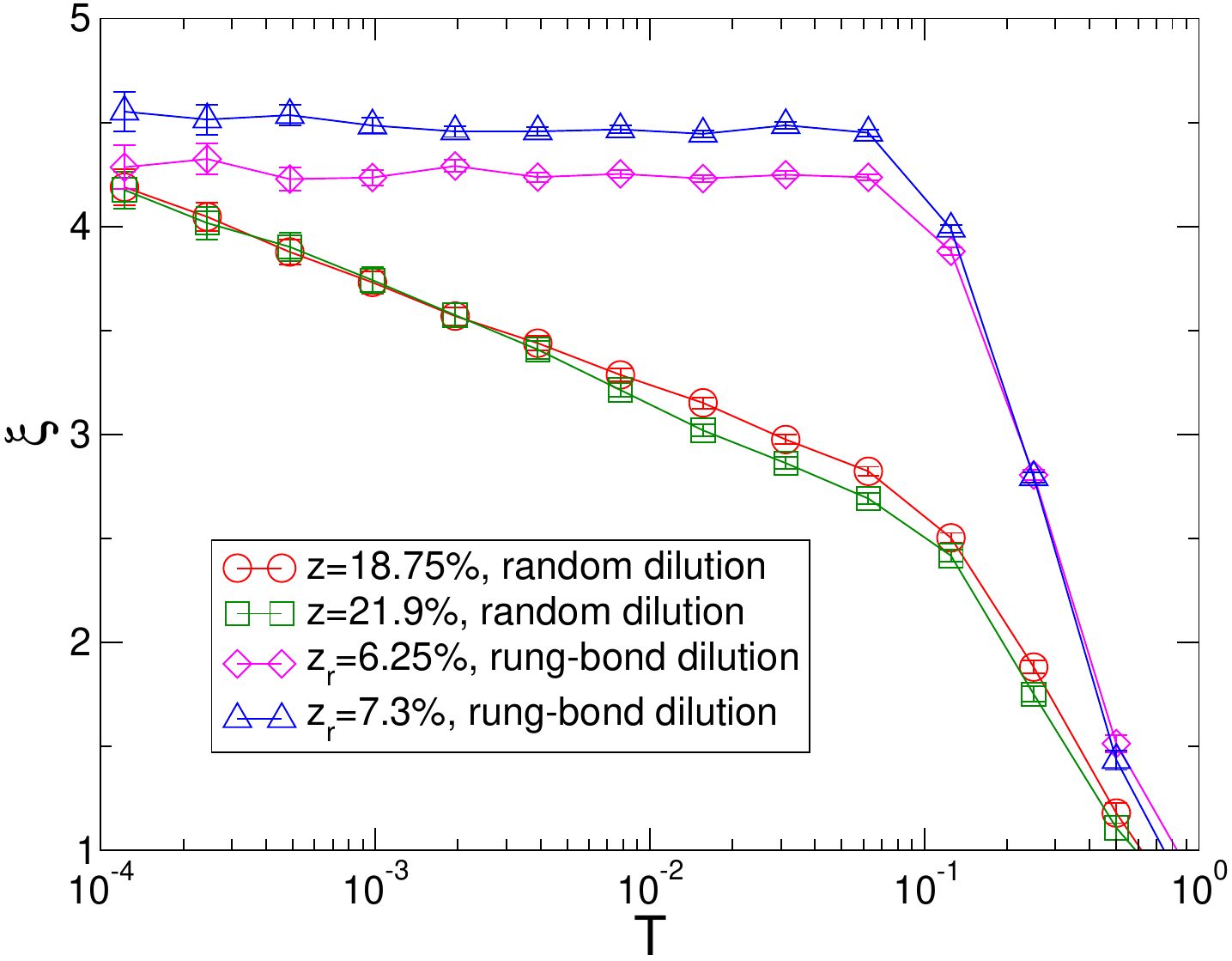}
\caption{\label{para.z} (color online). (a) Temperature dependence of correlation lengths for four cases of rung-bond and random dilution. }
\end{figure}

The dependence of the coefficients $A(z)$ and $B(z)$ on bond dilution is shown in Fig.~\ref{f.bonddilution.T}(c) and (d). While $A(z)$ decreases with increasing bond dilution, $B(z)$ increases. In particular $B(z)$ appears to vanish for $z \lesssim z^*$, and it seems to suggest that a phase transition occurs at finite $z = z^*$ from a phase with correlations converging to a finite value at low $T$ to a phase with correlations diverging logarithmically with decreasing temperature. Obviously the divergent behavior of correlation length only persists up to lengths of the order $\langle l \rangle$, corresponding to the characteristic length of ladder segments, but the exceedingly slow growth of $\xi$ with decreasing $T$ keeps it safely below $\langle l \rangle$ for all the temperatures and doping values we explored.

To gain a deeper understanding on the above results, we study the temperature dependence of rung-bond dilution and random dilution separately. We consider rung-bond dilution with concentrations $z_{r} = z/3$ such that the fraction of diluted rung bonds is the same as for the cases of random dilution that we have investigated. This allows us to quantitatively ascertain the separate effect of leg-bond and rung-bond dilution.  In presence of rung-bond dilution only, the correlation length is seen to converge to a finite value as $T \rightarrow 0$, even at relatively high concentration. Two examples with concentration $z_r=6.25\%$ and $7.3\%$ are shown in Fig.~\ref{para.z} (data for higher concentration, not shown here, exhibit a similar behavior). This implies that the system with rung-bond dilution only is gapless (because of the unbounded size of rare regions with \emph{e.g.} diluted adjacent rung bonds) and it has a finite correlation length, namely it is in a Griffiths-like phase. Adding leg-bond dilution (with concentrations $12\%$ and $14.6\%$ respectively), we recover a randomly bond-diluted ladder (with concentrations $18.75\%$ and $21.9\%$ respectively). For large enough dilution the correlation is found to grow logarithmically with decreasing temperature, as in Eq.~\eqref{e.xi-bl}, which seems atypical for a Griffiths-like phase. On the other hand, we observe that, for dilution $z_{r} = z/3$, the correlation length for the rung-bond diluted ladder with concentration $z_{r}$ is an upper bound to that of the randomly diluted ladder with concentration $z$, a fact which is somewhat intuitive, given that rung-bond dilution is the main mechanism leading to the enhancement of correlations. Therefore it is a priori unclear whether the logarithmic growth of the randomly diluted ladder persists to even lower temperatures than the ones explored here, because this would eventually lead the correlation length of the randomly diluted ladder to exceed that of the rung-bond diluted one.

We can then conclude that the correlation length results are \emph{a priori} consistent with two different scenarios for the randomly diluted 2-leg ladders. In a first scenario the system transitions from a Griffiths-like phase with a finite correlations for $z<z^*$ to a new phase with logarithmically diverging correlations, which is reminiscent of the behavior of a system controlled by an infinite-randomness fixed point (IRFP) -- see the discussion below. In a second scenario the correlation length exhibits different temperature dependences at intermediate temperatures for increasing bond dilution. Yet the correlation length of the randomly diluted ladders converges always to a finite value, because it is expected to be upper-bounded by that of rung-bond diluted ladders, which is seen to converge to a finite value. In this second scenario the system is therefore in a Griffiths-like phase for all the values of bond dilution we explored. In the following we find that the analysis of the temperature dependence of the uniform susceptibility helps clarifying which scenario is the most appropriate.

\subsubsection{Temperature dependence of the uniform susceptibility}

Fig.~\ref{xu.z}(a) shows the uniform susceptibility of a 2-leg ladder for various values of bond dilution. In the undoped system, the susceptibility of 2-leg ladders vanishes exponentially at low temperatures, in agreement with well-known previous results, and showing the gapped nature of the spectrum. In contrast, the susceptibility of the doped system diverges following a Curie law for low temperatures. Furthermore, Fig.~\ref{xu.z}(b) shows that the Curie coefficient increases as a function of bond dilution concentration following a power law, $C(z) \sim z^{2.66}$. One might expect that Curie paramagnetism comes trivially from spins which have remained isolated after bond dilution. Yet the concentration of such spins would scale as $z^3$ with dilution, given that one needs to dilute at least 3 bonds to decouple one spin from the rest of the ladder. Therefore the paramagnetism observed in the system has a collective nature. In fact, two contributions add to the one of free spins:

\begin{figure}
\includegraphics[width=8.0cm,angle=0]{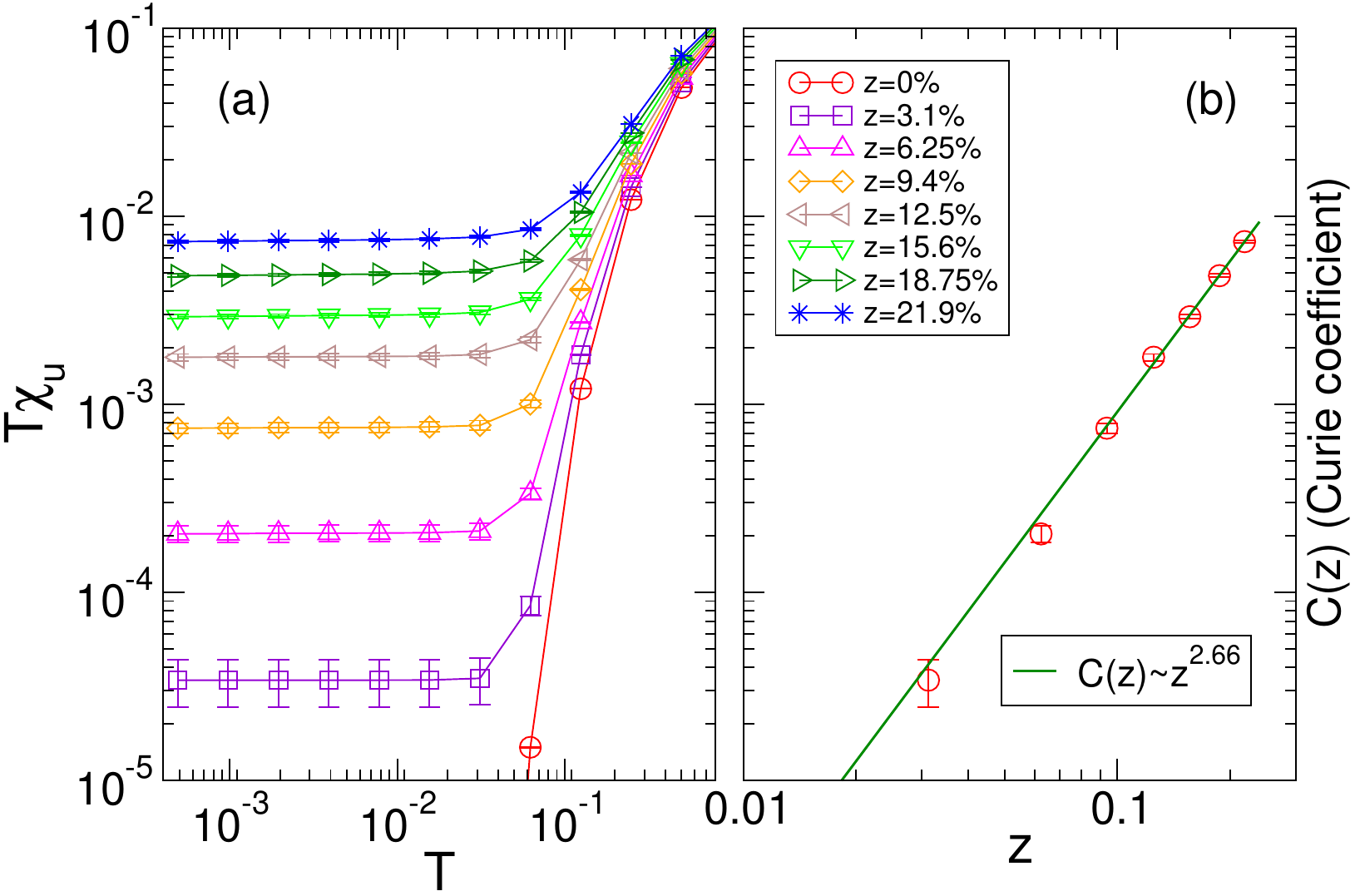}
\caption{\label{xu.z} (color online). (a) Low-temperature uniform susceptibility of bond-diluted ladders with $L=256$ and $J_l=J_r=1$. (b) The Curie coefficient increases as a power-law function of the bond dilution concentration $C(z) \sim z^{2.66}$.}
\end{figure}

\begin{itemize}
\item  odd-numbered clusters, obtained after fragmentation of the ladder into two (or more) pieces, have a doublet ground state, behaving as a collective spin 1/2. In our simulations on systems with an even total number of sites, three contiguous diluted bonds lead to an isolated site plus an odd-numbered cluster, so that the latter clusters have a probability $\sim z^3$. Alternatively odd-numbered clusters can be obtained by cutting the ladder at two distinct locations into two odd-numbered extended clusters (bigger than one single site). Yet one can easily see that this would require at least 5 diluted bonds at specific locations, giving rise to a probability $\sim z^5$, and consequently to a very small contribution to the uniform susceptibility. Therefore the contribution to collective paramagnetism cannot come from such clusters.
\item clusters with an even number of spins, but a non-equal number of spins on A and B sublattices, can also contribute to a diverging susceptibility. They represent therefore a further candidate for collective paramagnetism.
\end{itemize}

\begin{figure}
\includegraphics[width=8.0cm,angle=0]{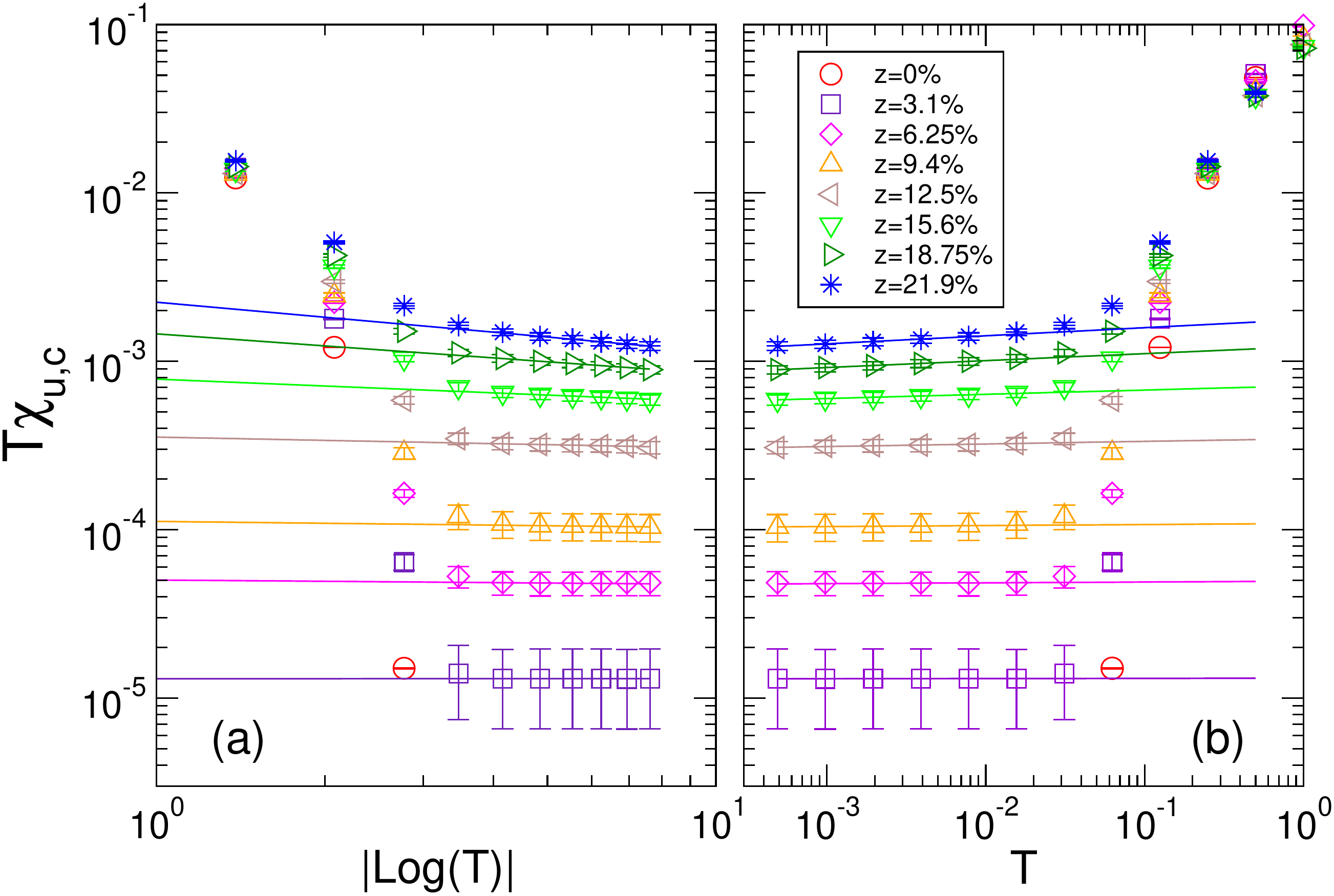}
\caption{\label{xuc.T} (color online). (a) Uniform susceptibility of even-numbered spin clusters of randomly bond-diluted 2-leg ladders. The solid lines are (a) logarithmic fits of the form $\chi_{u,c} = C/(T|\log T|^{\beta})$ and (b) power-law fits $\chi_{u,c} = C' T^{\beta'-1}$.}
\end{figure}

\begin{figure}
\includegraphics[width=8.0cm,angle=0]{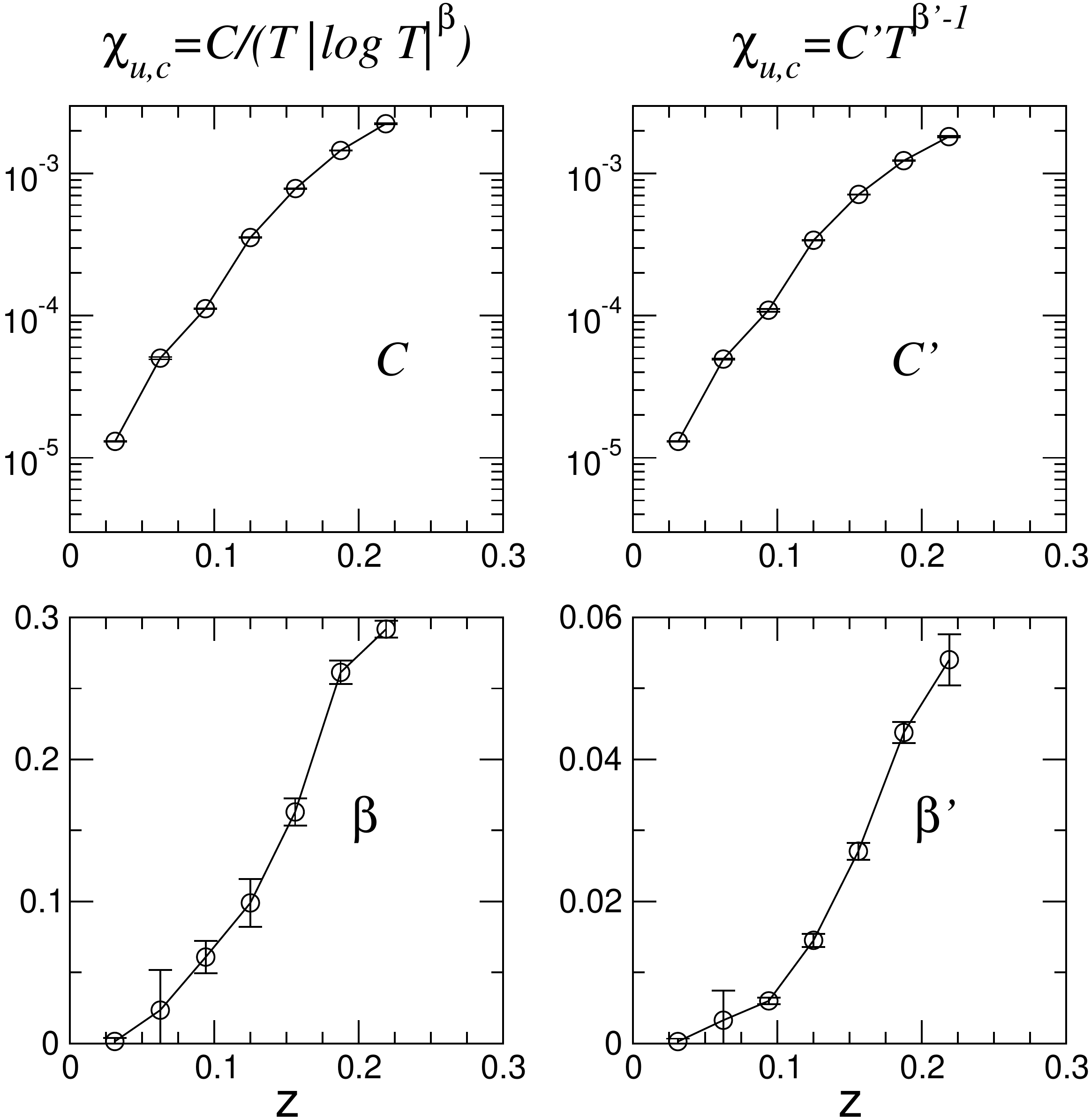}
\caption{\label{fit.para1} (color online). Fitting coefficients as a function of bond dilution. Left panels are fitting coefficients of the logarithmic function $\chi_{u,c} = C/(T|\log T|^{\beta})$. Right panels are fitting coefficients of the power-law function $\chi_{u,c} = C' T^{\beta'-1}$.}
\end{figure}

To isolate the second contribution, we calculate the uniform susceptibility coming from \emph{even-numbered} spin clusters only, $\chi_{u,c}$. This quantity is shown in Fig.~\ref{xuc.T}. We find that even-numbered uniform susceptibility at low temperatures can be equally well described by two fitting laws: a logarithmically-corrected Curie behavior:
\begin{equation}
\chi_{u,c} = \frac{C}{T |\log T|^{\beta}},
\end{equation}
and a power-law corrected Curie behavior:
\begin{equation}
\chi_{u,c} = \frac{C'}{T ^{1-\beta'}},
\label{e.Griffiths}
\end{equation}
In both cases we can conclude that the contribution to the uniform susceptibility coming from even-numbered clusters diverges more weakly than for a Curie law, and therefore that our system exhibits a highly non-trivial collective paramagnetism. The fitting coefficients $C, \beta$, and $C', \beta'$ for both cases are shown in Fig.~\ref{fit.para1}. We observe that all these fitting coefficients evolve smoothly with $z$, apparently contradicting the picture of a possible phase transition suggested by the behavior of the correlation length. In particular the non-universal power-law dependence of Eq.~\eqref{e.Griffiths} is typical of Griffiths behavior, as observed \emph{e.g.} in anisotropically bond-diluted square lattices \cite{Yuetal06}. Therefore the uniform susceptibility seems to favor the scenario for which the system is in a Griffiths phase for all values of the bond dilution we explored. For completeness we should also mention a third, less likely scenario, for which the correlation length is diverging logarithmically as $T\to 0$ even for infinitesimal doping (but the prefactor $B(z)$ of the logarithmic divergence becomes lower than our resolution). Therefore the system would be controlled by an IRFP for all doping values explored here; this scenario is somewhat consistent with the fact that a logarithmically corrected Curie law is consistent with the susceptibility at all doping values considered in this section, and that it could be another sign of IRFP behavior.

\subsubsection{Comparison with random ladders}

As sketched in Fig.~\ref{f.bonddil-cartoon}, the rung LMs in a bond-diluted ladder realize a ladder with weaker random couplings which are at once rung, leg and diagonal ones but without frustration. The rung couplings are randomized by the variety of different local environments which mediate the effective interaction between the rung LMs; while the leg couplings and the diagonal couplings are obviously randomized (in a correlated way) by the positional disorder of the diluted rung bonds.

For ladders with $J_l = J_r$ we cannot easily rely on perturbation theory to extract accurate expressions for the effective couplings, and an explicit calculation of the disorder statistics goes beyond the purpose of this paper. In any instance it is clear that the rung couplings obey a fundamentally different probability distribution with respect to the leg and the diagonal couplings: the rung couplings are always antiferromagnetic and weakly randomized, while the leg and diagonal couplings always take opposite signs and can be either ferromagnetic or antiferromagnetic - they are hence strongly disordered. To the best of our knowledge, such a peculiar model of a random ladder has not been investigated before.

In a critical quantum spin system the relationship between energy scales ($\sim \xi$) and length scales ($\sim T$ at finite temperature) is governed by the so-called dynamical critical exponent $\tilde z$, $\xi \sim T^{-1/{\tilde z}}$. The logarithmic dependence of the correlation length on temperature, Eq.~\eqref{e.xi-bl}, implies that  ${\tilde z}=\infty$: as mentioned before, this behavior, together with the logarithmically corrected Curie law in the susceptibility, could be suggestive of the fact that the bond-diluted ladder at sufficiently strong disorder is governed by an IRFP, similarly to what happens to \emph{e.g.} random antiferromagnetic $S=1/2$ chains \cite{Fisher94}. Nonetheless, it is not the same IRFP as in random antiferromagnetic chains, given that the correlation length and the susceptibility diverge differently (as  $\sim |\log (T/J)|^2$ and as $(T|\log T|^2)^{-1}$ respectively) in the latter. Existing RG studies of disordered ladder systems focus primarily on the case of antiferromagnetic ladders, either with nearest neighbor couplings only or with frustrated ones \cite{YusufY02, Melinetal02, HoyosM04, IgloiM05}. The occurrence of extended parameter regions with IRFP is only observed in the case of frustrated ladders. Therefore it is quite remarkable to observe an IRFP-like behavior in our unfrustrated system for a large interval of doping values. On the other hand, the absence of a documented IRFP for unfrustrated random ladders suggest that the logarithmic divergence observed in our system might not be the true asymptotic behavior for $T\to 0$. Nonetheless, a definite conclusion on such an asymptotic behavior would require a precise determination of the distribution of the effective couplings between LMs, which goes beyond the scopes of this paper.

\subsection{4-leg and 6-leg ladders}

Fig.~\ref{f.bonddil-fixT} shows that the dependence of the correlation length on bond dilution in 4-leg and 6-leg ladders exhibits a qualitatively different behavior compared to 2-leg ladders. Indeed for 4-leg and 6-leg ladders $\xi$ is initially \emph{decreasing} with increasing bond dilution. In the case $J_l=J_r=J$ the correlation length reaches a minimum for $z \approx 6.2\%$, and then starts growing with doping, up to a maximum attained for $z\approx 22\%$. In the following we will analyze this phenomenon in details.

\begin{figure}
\includegraphics[width=8cm,angle=0]{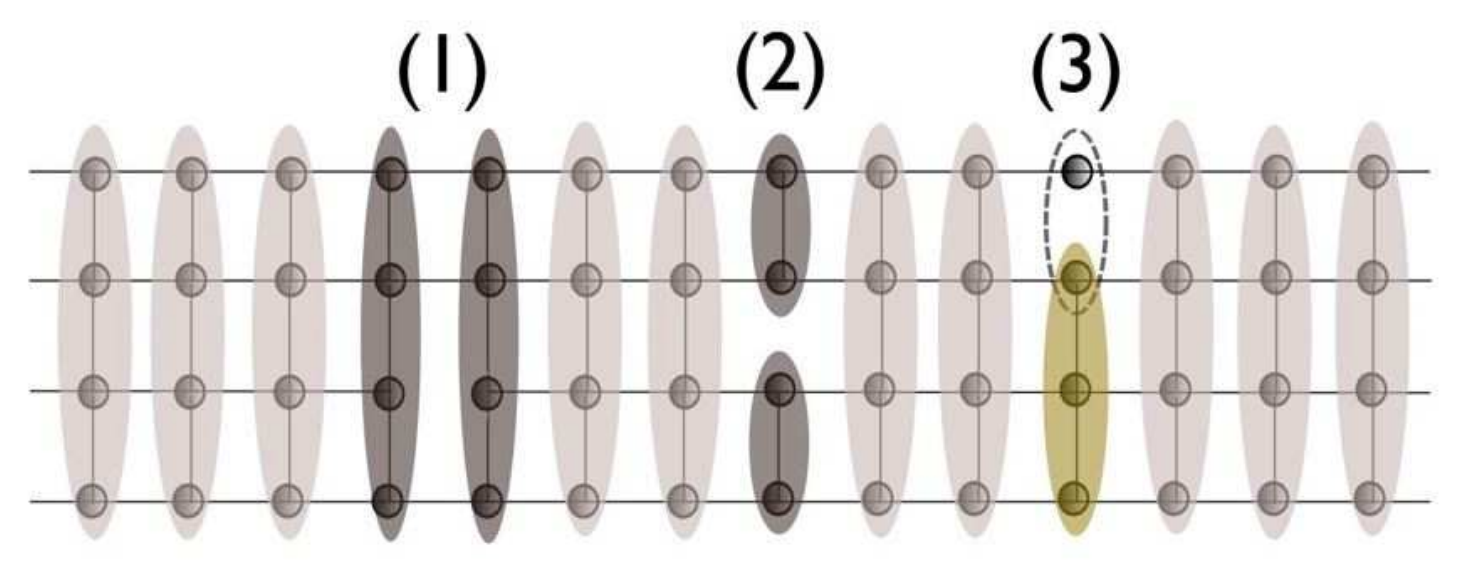}
\caption{\label{f.bonddil-cartoon-4leg} (color online). Bond-dilution effects on 4-leg ladders: (1) rung-singlet enhancement upon leg-bond dilution; (2) center-rung dilution leading to the formation of two 2-spin rung singlets; (3) outer-rung dilution with formation of a low-energy singlets between a LMs and a 3-spin doublet.}
\end{figure}

Similarly to what done for 2-leg ladders, we can distinguish among different effects of doping depending on which bond is diluted; we will focus here on the case of a 4-leg ladder (which is sketched in Fig.~\ref{f.bonddil-cartoon-4leg}):
\begin{enumerate}
\item the dilution of a \emph{leg bond} leads to the the enhancement of the 4-spin rung singlet on the two adjacent rungs;
\item the dilution of a \emph{center rung bond} leads to the formation of two 2-spin rung singlets on the intact outer rungs;
\item the dilution of an \emph{outer rung bond} separates an outer LM from three rung spins, which tend to form locally a spin-doublet. This doublet should have then tendency to form again a low-energy singlet with the LM through the effective antiferromagnetic coupling which binds them.
\end{enumerate}

Hence we observe that bond dilution of type 1) and 2) have the tendency to enhance locally the spin gap of the ladder, while dilution of type 3) has the tendency to suppress it. The case of 6-leg ladders is analogous, with the only difference that dilution of the central rung bond leads to the formation of two rung doublets. Therefore we will hereafter focus on the 4-leg case for simplicity.

We have disentangled the competing effects of dilution on 4-leg ladders by considering only one at a time. First of all we consider the effect of enhancement of correlations due to outer rung-bond dilution only, whose results are shown in Fig.~\ref{f.rungdilution}. Analogously to what has been seen for rung-bond dilution in the 2-leg ladders, outer-rung-bond dilution in 4-leg ladders leads to a strong enhancement of correlations (with a similar exponential scaling $\xi_{\rm or}(z) \approx \xi_0 \exp(a' z)$, or=outer rung).  In homogeneously diluted ladders this effect of enhancement is fundamentally limited by the finiteness of ladder segments (due to leg-bond dilution): indeed a numerical estimate of the average length of ladder segments, $\langle l \rangle$, shows that $\langle l \rangle$ crosses the correlation length $\xi_{\rm or}$ for a value of $z$ ($\approx 23\%$) which nicely corresponds to the optimal doping in the homogeneously bond-diluted ladder. Hence, similarly to 2-leg ladders, we can conclude that the optimal doping is completely dictated by the competing effect of outer-rung-bond dilution and ladder fragmentation.

\begin{figure}
\includegraphics[width=8.0cm,angle=0]{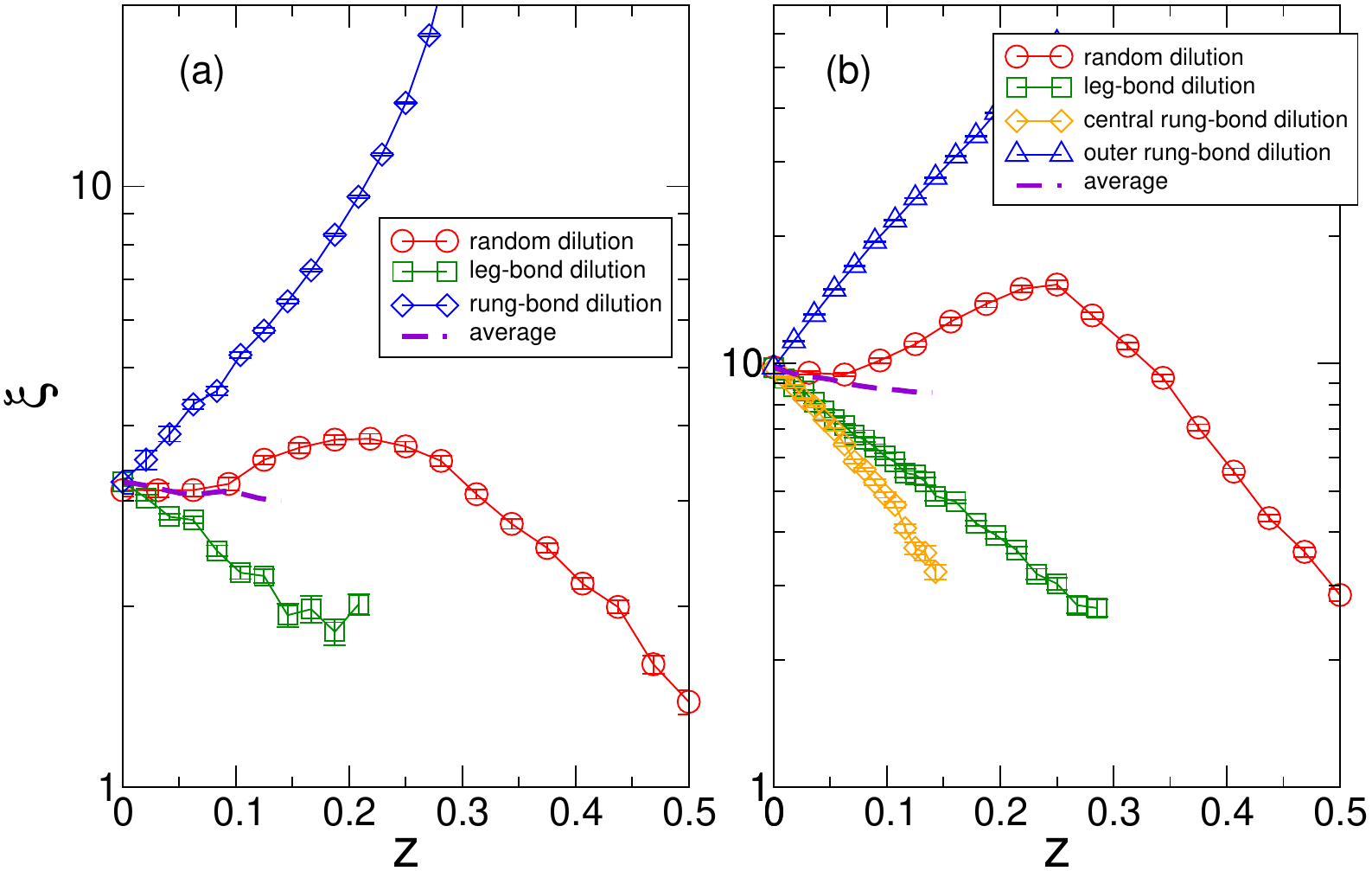}
\caption{\label{f.selective-dilution} (color online). Correlation length with selective bond dilution. Bond dilution is distinguished as: (a) rung-bond and leg-bond dilution in 2-leg ladders; (b) center-rung bond, outer-rung bond and leg bond dilution in 4-leg ladders. The dashed line is the average correlation length from above selectively bond-diluted mechanisms}
\end{figure}

Finally, we study the effect of selective dilution on the correlation length. In Fig.~\ref{f.selective-dilution}(a), we investigate the effect of rung-bond and leg-bond dilution in 2-leg ladders. In Fig.~\ref{f.selective-dilution}(b), we distinguish dilution of outer-rung bonds, center-rung bonds, and leg bonds in 4-leg ladders. As anticipated, latter forms of dilution in above cases lead to a significant suppression of correlations, again with approximate exponential scaling, $\xi_{\rm l,cr}(z) \approx \xi_0 \exp(-b' z)$ (l = leg, cr = center rung). We observe that  $b' \approx a'$ ($b \approx a$ for 2-led ladder), so that the simultaneous suppression and enhancement of correlations due to dilution of different types of bonds are in strong competition. The fact that at low doping the suppression of correlation wins is simply due to combinatorics: in 2-leg ladders (a), rung bonds only hold a fraction $f_r=1/3$ while leg bonds contain $f_l=2/3$ of all bonds; in 4-leg ladders (b), outer-rung bonds only represent a fraction $f_{\rm or}=2/7$ of all the bonds, whereas leg bonds and center-rung bonds are respectively $f_{\rm l} = 4/7$ and $f_{\rm cr}= 1/7$. Hence correlation-enhancing bond dilution has a global lower probability than correlation-suppressing one. In the limit of very low dilution ($z \ll 1/{\xi_0}$) we can easily imagine to separate a ladder into correlated regions (of characteristic length $\xi_0$) which are statistically independent from each other.  Upon dilution, each correlated region will only be affected by a single missing bond on average, with a probability $f_i$ for the $i$-th bond type ($i$ = r, l in 2-leg ladders; and $i$ = or, l, cr in 4-leg ladders).  Therefore the local correlations will be enhanced or suppressed in the same way as for the case of selective bond doping of type $i$. For $z \ll 1/{\xi_0}$  it is then tempting to write the correlation length of the homogeneously diluted ladder as a spatial average of different local correlation lengths. Assuming that the ratio of structure factors entering in Eq.~\eqref{e.2ndmoment} can be written as
 \begin{equation}
 \frac{[S(\pi,\pi)]_{\rm av}}{[S(\pi+2\pi/L,\pi)]_{\rm av}} \approx \sum_i f_i \frac{[S(\pi,\pi)]_{{\rm av},i}}{[S(\pi+2\pi/L,\pi)]_{{\rm av},i}}
 \end{equation}
 where $[...]_{{\rm av},i}$ denotes averaging over disorder realizations with dilution of bond of $i$-th type only (and with dilution concentration $f_i z$), we obtain that
 \begin{equation}
\xi(z) \approx \left[\sum_i f_i ~\xi_i^2(f_i z)\right]^{1/2}~.
 \end{equation}
 This average correlation length is presented by the dashed lines in Fig.~\ref{f.selective-dilution}. It clearly shows a quantitative agreement with random bond dilution results at small dilution.

\section{Conclusions}

In this paper we have discussed the effects of site and bond dilution on the low-temperature correlations of even-leg $S=1/2$ ladders with antiferromagnetic Heisenberg interactions. Site dilution is found to prune rung singlets and thus create localized moments which interact through unfrustrated, distance dependent couplings. The Hamiltonian describing the effective interaction between these moments is a random exchange Heisenberg model (REHM), with a gapless spectrum and power-law diverging correlations as temperature decreases to zero. We find that the distribution of the effective couplings has an intrinsically discrete structure, which prevents the system from realizing the universal regime of the REHM with continuous couplings. Further studies would hence be desired to clarify the role of a discrete distribution of effective couplings in determining the fixed point which governs the low temperature physics of this system.

Bond dilution, on the other hand, can either enhance or suppress locally the spin gap in even-leg ladders, depending on the location of the bond which is diluted. As a result of these competing effects, weak bond dilution is not enhancing the correlations of pure ladders. In fact, correlations can even be suppressed by disorder, especially in 4-leg and 6-leg ladders. The resulting short-range correlated phase has a gapless, Griffiths nature, due to the appearance of exponentially rare, but locally gapless regions. Beyond a critical concentration, the correlation-enhancing effect of bond dilution leads to a phase in 2-leg ladders, with a logarithmic scaling of the correlation length with decreasing temperature. This behavior may suggests that the low-temperature behavior of the system is governed by an infinite randomness fixed point over an extended range of parameters. Nonetheless, based on the investigation of the uniform susceptibility of the system, and by comparison with existing results on disordered ladders, we argue that this might not be the true asymptotic behavior for $T\to 0$, and that the system is more likely to remain in a Griffiths phase for all values of bond dilution.

In order to connect our results to experiments on site-diluted and bond-disordered ladders, a fundamental question to address is the role of finite inter-ladder couplings, which are unavoidable in real materials. If $J'$ is their energy scale, such couplings drive the system towards a three-dimensional magnetically ordered phase below a temperature $T_c$, which can be estimated via a mean-field approach as $k_B T_c \approx J'~ \xi(T_c;z)$, where $\xi$ is the correlation length of the uncoupled ladders. For a gapless critical phase, as that induced by site dilution (at any concentration) and by bond dilution (beyond a critical concentration), $\xi(T;z)$ diverges for $T\to 0$, so that the system orders at finite temperature. On the other hand, the divergence of $\xi$ with decreasing temperature appears to be relatively weak (power law with a small power $\sim 0.3-0.4$ for site dilution, or logarithmic for bond dilution). Hence a difference of several orders of magnitude between the intra-ladder and the interladder couplings, which is common to various ladder materials, allows the asymptotic low-temperature behavior of $\xi$ for the uncoupled doped ladders to be manifested at temperatures which lie well above the critical temperature $T_c$ for magnetic ordering. For bond-diluted ladders, moreover, short-range correlations proper of the pure ladder persist up to a critical doping, so that three-dimensional ordering in coupled ladders should be absent at weak dilution. Increasing the bond dilution leads to an increase of correlations at low temperature, so that the system is expected to undergo a disorder-induced quantum phase transition from a Griffiths phase to a magnetically ordered phase (at $T=0$), but with exponentially small transition temperatures, $T_c\sim \exp(-J/J')$, which might be hardly detectable in experiments. As a result the temperature scaling of correlations investigated here for uncoupled ladders is expected to be relevant to the behavior of real ladder materials with non-magnetic doping.

\begin{acknowledgments}
We would like to thank N. Bray-Ali, S. Garnerone, and B. Normand for valuable discussions. We also acknowledge financial support by the Department of Energy, grant number DE-FG03-01ER45908. T. R. acknowledges the hospitality of the MPIPKS-Dresden, where part of this work was completed. R. Y. was supported by NSF Grant No. DMR-1006985 and Robert A. Welch Foundation Grant No. C-1411. The numerical computations were carried out on the University of Southern California high-performance supercomputer cluster.
\end{acknowledgments}


\begin{thebibliography}{200}
\bibitem{DagottoR96} E. Dagotto, T. M. Rice, Science {\bf 271}, 618 (1996).
\bibitem{SigristF96} M. Sigrist and A. Furusaki, J. Phys. Soc. Jpn. {\bf 65}, 2385 (1996).
\bibitem{Sandviketal97} A.W. Sandvik, E. Dagotto, and D. J. Scalapino, Phys. Rev. B {\bf 56}, 11701 (1997).
\bibitem{Martinsetal97} G.B. Martins, M. Laukamp, J. Riera, and E. Dagotto, Phys. Rev. Lett. {\bf 78}, 3563 (1997).
%\bibitem{Normand02} B. Normand and F. Mila, Phys. Rev. B 65, 104411 (2002).
\bibitem{Wessel01} S. Wessel, B. Normand, M. Sigrist, and S. Haas, Phys. Rev. Lett. {\bf 86}, 1086 (2001).
\bibitem{Laflorencie03} N. Laflorencie and D. Poilblanc, Phys. Rev. Lett. {\bf 90}, 157202 (2003).
\bibitem{Yasuda01} C. Yasuda, S. Todo, M. Matsumoto, and H. Takayama, Phys. Rev. B {\bf 64}, 092405 (2001).
\bibitem{Imada97} M. Imada and Y. Iino, J. Phys. Soc. Jpn. {\bf 66}, 568 (1997).
\bibitem{Weber06} H. Weber and M. Vojta, Eur. Phys. J. B {\bf 53}, 185 (2006).
\bibitem{Azumaetal97} M. Azuma, Y. Fujishiro, M. Takano, M. Nohara and H. Takagi, Phys. Rev. B. {\bf 55}, R8658 (1997).
\bibitem{Bobroffetal09} J. Bobroff, N. Laflorencie, L. K. Alexander, A. V. Mahajan, B. Koteswararao, and P. Mendels, Phys. Rev. Lett. {\bf 103}, 047201 (2009).
\bibitem{Manakaetal08} H. Manaka, A. V. Kolomiets, and T. Goto, Phys. Rev. Lett. {\bf 101} 077204 (2008).
\bibitem{Manakaetal09} H. Manaka, H. A. Katori, O. V. Kolomiets, and T. Goto, Phys. Rev. B {\bf 79} 092401 (2009).
\bibitem{Hongetal10} Tao Hong, A. Zheludev, H. Manaka, and L.-P. Regnault, Phys. Rev. B {\bf 81} 060410 (2010).
\bibitem{GrevenB98} M. Greven and R. J. Birgeneau, Phys. Rev. Lett. {\bf 81}, 1945 (1998).
\bibitem{Miyazakietal97} T. Miyazaki, M. Troyer, M. Ogata, K. Ueda, and D. Yoshioka, J. Phys. Soc. Jpn. {\bf 66}, 2580 (1997).
\bibitem{Furusakietal94} A. Furusaki, M. Sigrist, P. A. Lee, K. Tanaka, and N. Nagaosa, Phys. Rev. Lett. {\bf 73}, 2622 (1994).
\bibitem{Westerbergetal95} E. Westerberg, A. Furusaki, M. Sigrist, and P. A. Lee, Phys. Rev. Lett. {\bf 75}, 4302 (1995).
\bibitem{Nagaosaetal96} N. Nagaosa, A. Furusaki, M. Sigrist, and H. Fukuyama,  J. Phys. Soc. Jpn. {\bf 67}, 3724 (1996).
\bibitem{Yuetal05} R. Yu, T. Roscilde, and S. Haas, Phys. Rev. Lett. {\bf 94}, 197204 (2005).
\bibitem{Yuetal06} R. Yu, T. Roscilde, and S. Haas, Phys. Rev. B {\bf 73}, 064406 (2006).
\bibitem{Yasuda06}  C. Yasuda, S. Todo, and H Takayama,  J. Phys. Soc. Jpn. {\bf 75}, 124704 (2006).
\bibitem{SyljuasenS02} O. F. Sylju\aa sen and A. W. Sandvik, Phys. Rev. E {\bf 66}, 046701 (2002).
\bibitem{Sandvik02} A. W. Sandvik, Phys. Rev. B {\bf 66} 024418 (2002).
\bibitem{Cooperetal82} F. Cooper, B. Freedman, and D. Preston, Nucl. Phys. B {\bf 210}, 210 (1982).
\bibitem{Mikeskaetal97} H.-J. Mikeska, U. Neugebauer, and U. Schollw\"ock, Phys. Rev. B {\bf 55}, 2955 (1997).
\bibitem{Westerbergetal97} E. Westerberg, A. Furusaki, M. Sigrist, and P. A. Lee, Phys. Rev. B {\bf 55}, 12578 (1997).
\bibitem{Frischmuthetal99} B. Frischmuth, M. Sigrist, B. Ammon, and M. Troyer, Phys. Rev. B {\bf 60}, 3388 (1999).
\bibitem{Laflorencie05} N. Laflorencie, D. Poilblanc, and M. Sigrist, Phys. Rev. B. {\bf 71}, 212403 (2005).
\bibitem{noteonlocalgap} The notion of local gap is associated to the local susceptibility at site $i$,
$\chi_{i} = (1/\beta) \int_0^{\beta} ~d\tau~\langle {\bm S}_i(\tau) {\bm S}_i(0) \rangle$. Assuming that
$\langle {\bm S}_i(\tau) {\bm S}_i(0) \rangle \sim \exp(-\Delta_i \tau)$ one obtains that $\chi_{i} \sim \Delta_i^{-1}$,
where $\Delta_i^{-1}$ is the local gap.
\bibitem{Stauffer85} D. Stauffer, "Introduction to Percolation Theory", Taylor and Francis, London 1985.
\bibitem{Newman} M. E. J. Newman and R. M. Ziff, Phys. Rev. Lett. {\bf 85}, 4104 (2000); M. E. J. Newman and R. M. Ziff, Phys. Rev. E {\bf 64}, 016706 (2001).
\bibitem{Fisher94} D. S. Fisher, Phys. Rev. B {\bf 50}, 3799 (1994).
\bibitem{YusufY02} E. Yusuf and K. Yang, Phys. Rev. B {\bf 65}, 224428 (2002).
\bibitem{Melinetal02} R. M\'elin, Y.-C. Lin, P. Lajk\'o, H. Rieger, and F. Igl\'oi, Phys. Rev. B {\bf 65}, 104415 (2002).
\bibitem{HoyosM04} J. A. Hoyos and E. Miranda, Phys. Rev. B {\bf 69}, 214411 (2004).
\bibitem{IgloiM05} F. Igl\'oi and C. Monthus, Phys. Rep. {\bf 412}, 277 (2005).
\bibitem{footnote1} It is to be said that, at the low temperature to which the figure refers ($T =  J_l /1024$), the correlation length on the doped 4-leg and 6-leg ladders grows to values which are no longer satisfying the condition $\xi(T) \ll L$, and hence it is limited by the system size ($L = 128$) explored there.
\end{thebibliography}
\end{document}